\documentclass[pdflatex,sn-mathphys,Numbered]{sn-jnl}

\usepackage{graphicx}%
\usepackage{multirow}%
\usepackage{amsmath,amssymb,amsfonts}%
\usepackage{amsthm}%
\usepackage{mathrsfs}%
\usepackage[title]{appendix}%
\usepackage{xcolor}%
\usepackage{textcomp}%
\usepackage{manyfoot}%
\usepackage{booktabs}%
\usepackage{algorithm}%
\usepackage{algorithmicx}%
\usepackage{algpseudocode}%
\usepackage{listings}%

\usepackage{verbatim}
\usepackage[version=4]{mhchem}
\usepackage{physics}
\usepackage{placeins}
\usepackage[absolute,overlay]{textpos}
\usepackage{caption}
\usepackage{subcaption}
\usepackage{expl3}
\usepackage{pgffor}
\usepackage{cleveref}
\usepackage{soul}
\usepackage{floatpag}
\makeatletter
\def\ps@plain{%
  \let\@mkboth\@gobbletwo
  \let\@oddhead\@empty
  \let\@evenhead\@empty
  \def\@oddfoot{\vbox to 18pt{\vfill\reset@font\rmfamily\hfil\thepage\hfil}}%
  \let\@evenfoot\@oddfoot
}
\makeatother

\usepackage{silence}
\renewcommand{\thesubsection}{\arabic{subsection}}

\begin{document}

\title[Article Title]{Synthesis-route-dependent~state~selection in~the~frustrated~spinel~\ce{ZnV2O4}}

\author[1]{\fnm{Jem} \sur{Pitcairn}}
\author[1,2]{\fnm{Jennifer N.} \sur{Graham}}
\author[1,3]{\fnm{Jacob L.} \sur{Brownlee}}
\author[1]{\fnm{Shruti} \sur{Chakravarty}}
\author[4,5]{\fnm{Joya A.} \sur{Cooley}}
\author[6,7]{\fnm{Giorgia} \sur{Confalonieri}}
\author[3]{\fnm{A. Dominic} \sur{Fortes}}
\author[2]{\fnm{David W.} \sur{Tam}}
\author*[3]{\fnm{J. Ross} \sur{Stewart}}\email{l.m.clark@bham.ac.uk}
\author*[1]{\fnm{Lucy} \sur{Clark}}\email{ross.stewart@stfc.ac.uk}

\affil[1]{\orgdiv{School of Chemistry}, \orgname{University of Birmingham}, \orgaddress{\city{Edgbaston, Birmingham}, \postcode{B15~2TT}, \country{UK}}}
\affil[2]{\orgname{Institut Laue-Langevin}, \orgaddress{\street{71 Avenue des Martyrs}, \city{38042 Grenoble}, \country{France}}}
\affil[3]{\orgdiv{ISIS Neutron and Muon Source}, \orgname{Rutherford Appleton Laboratory}, \orgaddress{\street{Harwell Campus}, \city{Didcot}, \postcode{OX11~0QX}, \country{UK}}}
\affil[4]{\orgdiv{Materials Research Laboratory}, \orgname{University of California Santa Barbara}, \orgaddress{\city{Santa Barbara}, \state{California 93106}, \country{USA}}}
\affil[5]{\orgdiv{Department of Chemistry and Biochemistry}, \orgname{California State University, Fullerton}, \orgaddress{\city{Fullerton}, \state{California 92831}, \country{USA}}}
\affil[6]{\orgname{European Synchrotron Research Facility}, \orgaddress{\street{71~Avenue~des~Martyrs}, \city{38043~Grenoble}, \country{France}}}
\affil[7]{\orgname{Department of Earth Sciences}, \orgname{Sapienza University of Rome}, \orgaddress{\street{Piazzale Aldo Moro 5}, \city{00185~Roma}, \country{Italy}}}

\abstract{The emergent properties of strongly correlated materials can be highly sensitive to how they are made, yet synthesis pathway is rarely treated as a fundamental variable governing their ground states. The spinel \ce{ZnV2O4} provides a model system in which to test this idea: its pyrochlore lattice of antiferromagnetically coupled V$^{3+}$ $S=1$ moments hosts competing spin and orbital degeneracies, while different synthesis routes yield markedly different low-temperature behaviour. Here, we show that changing the solid-state synthesis pathway reproducibly selects a distinct structural and magnetic ground state. Microwave-assisted solid-state synthesis (MASS) stabilises the cubic spinel structure to low temperature, suppressing the conventional cubic-to-tetragonal transition and preserving the highly frustrated $S=1$ pyrochlore lattice. Combining high-resolution diffraction, magnetometry and inelastic neutron scattering, we find a magnetic ground state exhibiting persistent spin-liquid-like dynamics together with partial spin freezing. Our results establish synthesis pathway as an experimental variable for selecting between competing emergent states in strongly correlated materials, revealing a route to accessing ground states not realised by conventional synthesis.}

\maketitle

\noindent The emergent properties of strongly correlated materials arise from complex interactions between their spin, orbital, charge and lattice degrees of freedom, giving rise to a wide range of collective quantum phenomena \cite{goyalExploringQuantumMaterials2025}. While considerable effort has focused on discovering and understanding these emergent states, comparatively little is understood about how synthesis pathway influences which ground state is ultimately realised \cite{Clark2026}. For example, this synthesis pathway dependence presents a particular challenge for candidate quantum spin liquids (QSLs), where competing interactions make magnetic behaviour highly sensitive to subtle variations in local structure, defects and disorder. QSLs are exotic magnetic states in which correlated quantum spin fluctuations persist at temperatures far below the energy scale set by the underlying magnetic exchange interactions \cite{broholmQuantumSpinLiquids2020}. Theoretically, they exhibit no long-range magnetic order, instead forming dynamically disordered, highly entangled ground states that can host fractional magnetic excitations \cite{broholmQuantumSpinLiquids2020,savaryQuantumSpinLiquids2017,zhouQuantumSpinLiquid2017}. These properties have driven an extensive search for experimental QSL candidates, including because of their potential relevance to quantum computation \cite{hwangAnyonCondensationConfinement2024}. Yet despite many theoretical model QSLs \cite{savaryQuantumSpinLiquids2017,zhouQuantumSpinLiquid2017,knolleFieldGuideSpin2019}, there remains no uncontentious experimental realisation across the many classes of candidate materials investigated \cite{takagiConceptRealizationKitaev2019,chamorroChemistryQuantumSpin2021,clarkQuantumSpinLiquids2021,goyalExploringQuantumMaterials2025}. \\

\noindent \noindent Experimentally, unambiguous identification of a QSL ground state is complicated by the sensitivity of competing magnetic interactions to structural and chemical disorder introduced during materials synthesis, leading to pronounced sample-to-sample variability \cite{bowmanRoleDefectsDetermining2019,arpinoImpactStoichiometryYb2Ti2O72017,tustainMagneticOrderQuantum2020}. The role of disorder itself is particularly difficult to interpret: it can induce spin freezing \cite{huFreezingDisorderInduced2021,martinDisorderQuantumSpin2017,gingrasQuantumSpinIce2014}, mimic QSL dynamics \cite{gingrasQuantumSpinIce2014,zhuDisorderInducedMimicrySpin2017,tustainMagneticOrderQuantum2020,savaryDisorderInducedQuantumSpin2017,furukawaQuantumSpinLiquid2015}, enhance quantum entanglement \cite{savaryDisorderInducedQuantumSpin2017,furukawaQuantumSpinLiquid2015,liuEvolutionQuantumEntanglement2017,chenAnomalousSuppressionLargescale2025}, and, recently, suppress the so-called eminuscent phase hidden in the ``disorder-free'' limit of QSL hosts \cite{syzranovEminuscentPhaseFrustrated2022}. Distinguishing intrinsic ground-state physics from synthesis-dependent behaviour therefore requires understanding how the pathway by which a material is prepared controls its structure, disorder and emergent properties. \\

\noindent Among strongly correlated materials, the spinel \ce{ZnV2O4} provides a particularly compelling model system in which to investigate synthesis-dependent emergent behaviour. At $T=300$~K, \ce{ZnV2O4} crystallises in the cubic spinel structure (Fig.~\ref{fig:cub}a), in which the V$^{3+}$ $3d^2$ $S=1$ ions form a pyrochlore network of corner-sharing tetrahedra (Fig.~\ref{fig:cub}b). Strong antiferromagnetic interactions between nearest-neighbour V$^{3+}$ moments cannot be satisfied simultaneously on this geometry, giving rise to pronounced geometric frustration \cite{Greedan2001}. In addition, the two $3d$ electrons of each V$^{3+}$ ion occupy the triply degenerate $t_{2g}$ manifold of the surrounding octahedral crystal field, introducing an additional orbital degree of freedom. Previous estimates place the energy scale for antiferromagnetic exchange coupling in \ce{ZnV2O4} in the range of $400-1000$~K \cite{uedaMagneticStructuralTransitions1997,reehuisCrystallographicMagneticStructure2003,leeOrbitalSpinChains2004,vasilievSpecificHeatMagnetic2006,wangPhaseTransitionCritical2015,browneOrbitalMoleculesVanadium2020,singhaStructuralMagneticStudies2019}. The resulting interplay of spin, orbital and lattice degrees of freedom makes \ce{ZnV2O4} a long-standing model system for exploring unconventional magnetic ground states. While theoretical studies of QSLs have historically focused on frustrated $S=1/2$ systems \cite{canalsPyrochloreAntiferromagnetThreeDimensional1998,hermelePyrochlorePhotonsU12004,schaferPyrochlore122020,hallasXYAntiferromagneticGround2016}, recent work has highlighted the rich phase space accessible within the $S=1$ pyrochlore antiferromagnet \cite{hagymasiEnhancedSymmetrybreakingTendencies2022,gaoSpin1PyrochloreAntiferromagnets2020,iqbalQuantumClassicalPhases2019,mullerThermodynamicsPyrochlorelatticeQuantum2019}, including several QSL-like phases \cite{pohleAbundanceSpinLiquids2025}. Experimentally, the absence of magnetic long-range order above at least $T=40$~K, despite the large antiferromagnetic exchange energy scale \cite{reehuisCrystallographicMagneticStructure2003}, highlights \ce{ZnV2O4} as a promising system in which to explore highly frustrated magnetic ground states. However, this potential has remained largely unrealised because of the pronounced and unresolved synthesis dependence of its structural and magnetic properties. \\

\noindent The most widely studied polycrystalline \ce{ZnV2O4} samples, prepared \textit{via} conventional furnace-heated solid-state synthesis, typically undergo a cubic-to-tetragonal structural transition at $T_\mathrm{S}\approx50$~K \cite{uedaMagneticStructuralTransitions1997,reehuisCrystallographicMagneticStructure2003,leeOrbitalSpinChains2004,wangPhaseTransitionCritical2015,browneOrbitalMoleculesVanadium2020}. This is thought to be driven by a local Jahn-Teller distortion of the V$^{3+}$ $3d^2$ octahedral crystal field and subsequent orbital ordering \cite{motomeOrbitalMagneticTransitions2004,tchernyshyovStructuralOrbitalMagnetic2004,browneOrbitalMoleculesVanadium2020,reehuisCrystallographicMagneticStructure2003,singhaStructuralMagneticStudies2019}. A spin-ordering transition closely follows $T_\mathrm{S}$ and is proposed to mark the onset of long-range antiferromagnetic order at $T_\mathrm{N}\approx40$~K \cite{uedaMagneticStructuralTransitions1997,reehuisCrystallographicMagneticStructure2003,leeOrbitalSpinChains2004,wangPhaseTransitionCritical2015,browneOrbitalMoleculesVanadium2020,singhaStructuralMagneticStudies2019,blanco-canosaEnhancedPressureDependence2007}. However, low-temperature neutron diffraction data showing magnetic Bragg peaks associated with this antiferromagnetic order have only been reported for two samples, and the observed reflections do not appear to be resolution limited \cite{reehuisCrystallographicMagneticStructure2003,niziolInvestigationMagneticProperties1973}. Furthermore, the transition temperatures themselves are highly sample dependent (Table~\ref{tab:ZVO_Ts}, Supplementary Information), and there remains no consensus regarding the structural, orbital and magnetic ground states of \ce{ZnV2O4}, either experimentally or theoretically \cite{motomeOrbitalMagneticTransitions2004,tchernyshyovStructuralOrbitalMagnetic2004,browneOrbitalMoleculesVanadium2020,leeOrbitalSpinChains2004,singhaStructuralMagneticStudies2019,pardoHomopolarBondFormation2008,khomskiiOrbitallyInducedPeierls2005,maitraOrbitalOrderZnV2O42007,dimatteoOrbitalOrderVanadium2005}. \\

\noindent The influence of synthesis pathway is perhaps most clearly illustrated by comparison with single crystals obtained \textit{via} chemical vapour transport \cite{ebbinghausCrystalStructureMagnetic2004} and nanoparticles prepared \textit{via} sol--gel synthesis \cite{shrivastavaModulatingOpticalMagnetic2020}. Unlike furnace-prepared polycrystalline samples, neither exhibits evidence for the characteristic structural and magnetic transitions near 50 and 40~K, respectively \cite{ebbinghausCrystalStructureMagnetic2004,shrivastavaModulatingOpticalMagnetic2020}. Instead, both studies suggest the emergence of spin freezing at $T_\mathrm{f}\approx11$~K. Unlike long-range antiferromagnetic order, spin freezing describes the slowing of spin fluctuations towards a static disordered magnetic state without the development of long-range magnetic order. Collectively, these observations demonstrate that the structural and magnetic properties of \ce{ZnV2O4} are highly sensitive to how the material is prepared. However, the low-temperature states produced by these alternative synthesis routes have not been comprehensively characterised using complementary high-resolution diffraction and spectroscopic measurements. Consequently, the relationship between synthesis pathway and the resulting structural and magnetic behaviour remains poorly understood. \\

\noindent To address this, we developed a microwave-assisted solid-state synthesis (MASS) route to \ce{ZnV2O4} \cite{levinProtocolsHighTemperature2019}. The distinct heating mechanism and substantially shortened reaction times associated with MASS reproducibly yield highly crystalline polycrystalline samples with markedly different structure-property behaviour from conventional furnace synthesis. Through complementary high-resolution diffraction, magnetometry and inelastic neutron scattering, we show that MASS suppresses the conventional cubic-to-tetragonal transition and stabilises the orbitally degenerate cubic spinel structure to low temperature. The resulting magnetic ground state exhibits short-range correlated spin-liquid-like dynamics together with partial spin freezing. These results establish synthesis pathway as an experimental variable for selecting between competing emergent states in \ce{ZnV2O4}, demonstrating that fundamentally different ground states can be accessed without changing the nominal chemical composition.

\section*{Synthesis-pathway stabilisation of the cubic crystallographic ground state}

\noindent MASS provides access to heating conditions and reaction timescales markedly different from conventional furnace synthesis and has been applied to the rapid preparation of a wide range of complex inorganic materials \cite{levinProtocolsHighTemperature2019,Ortiz2022,Lekse2007}. Substantially reduced reaction times and energy consumption can be achieved while maintaining high sample crystallinity \cite{Clemen2025}, providing a distinct synthesis pathway through which to investigate synthesis-dependent behaviour in strongly correlated materials. Our MASS method uses the same starting materials as the conventional furnace-based synthesis widely reported for polycrystalline \ce{ZnV2O4} \cite{uedaMagneticStructuralTransitions1997,reehuisCrystallographicMagneticStructure2003,leeOrbitalSpinChains2004,vasilievSpecificHeatMagnetic2006,wangPhaseTransitionCritical2015,browneOrbitalMoleculesVanadium2020,singhaStructuralMagneticStudies2019}. Whereas furnace-based synthesis requires heating to $T\geq1073$~K for up to 17 days \cite{reehuisCrystallographicMagneticStructure2003}, the direct and localised heating provided by the microwave susceptor reduces the synthesis time to under 15 minutes (see Methods). This three-order-of-magnitude reduction in reaction timescale makes possible the rapid optimisation of synthesis conditions and the preparation of the large sample masses required for inelastic neutron scattering measurements. Moreover, the substantially shorter exposure to high-temperature reaction conditions may help to mitigate synthetic challenges associated with the volatility of the \ce{ZnO} precursor, which may lead to sample-dependent variations in stoichiometry, and the stability of the V$^{3+}$ oxidation state \cite{nohairCationicDistributionMechanism1994,katsuraEquilibriaV2O3VO2System1967,lideCRCHandbookChemistry2005}. We find that MASS provides a highly reproducible synthesis route to polycrystalline \ce{ZnV2O4}, with reproducibility of both the diffraction and magnetic susceptibility data demonstrated across independently prepared samples (Figs.~\ref{fig:synth_dep_2} and \ref{fig:synth_dep_3}, Supplementary Information). Whether this fundamentally different synthesis pathway also selects a different structural and magnetic ground state in \ce{ZnV2O4}, however, cannot be predicted \textit{a priori}. \\

\begin{figure}[hbt!]
    \centering
    \includegraphics{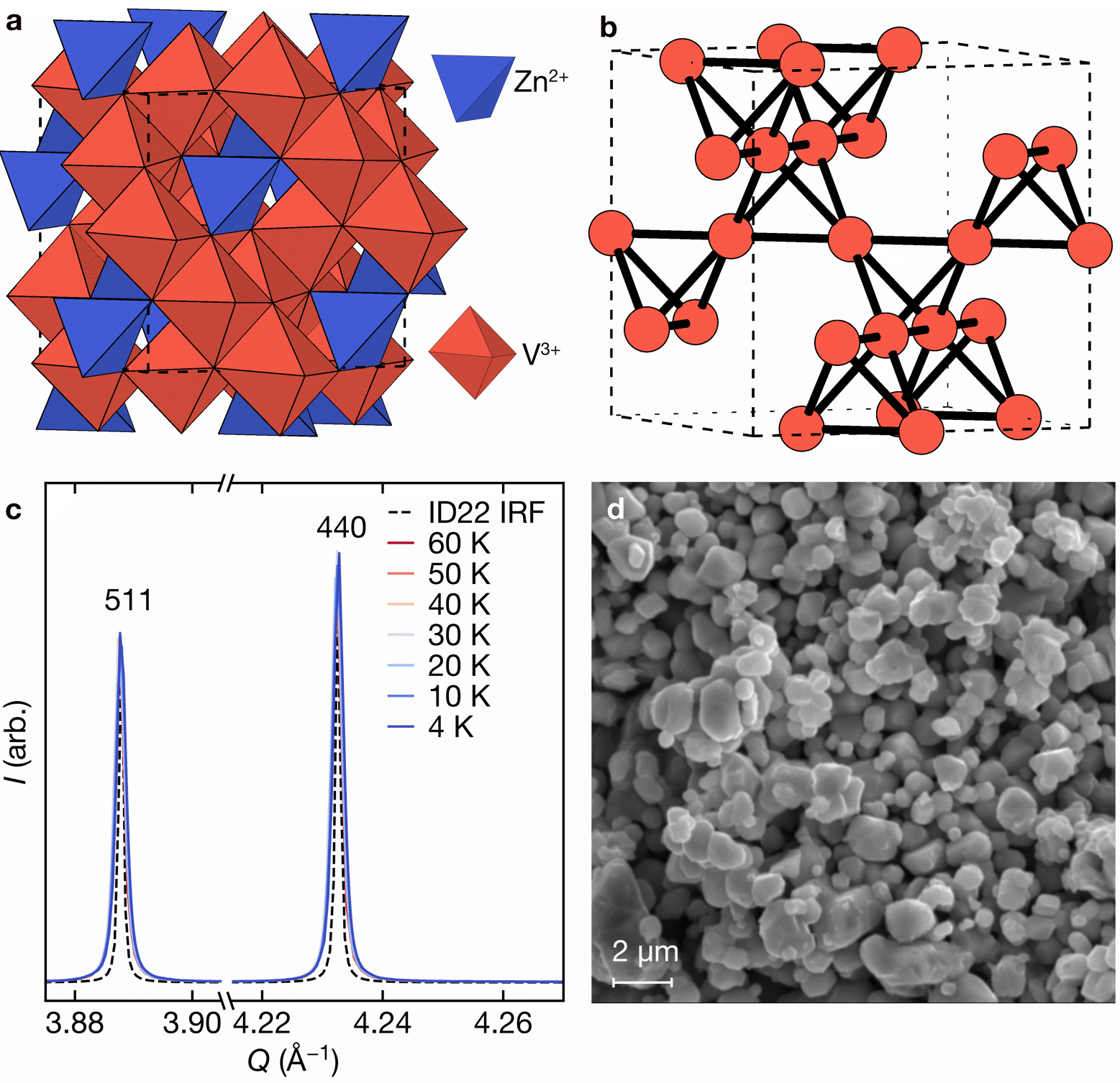}
    \caption{\textbf{Crystal structure and morphology of MASS-produced spinel \ce{ZnV2O4}.} \textbf{a} Crystal structure of cubic spinel \ce{ZnV2O4} ($Fd\bar{3}m$), in which \ce{Zn^{2+}} and \ce{V^{3+}} ions occupy tetrahedral (blue polyhedra) and octahedral (red polyhedra) sites, respectively. \ce{O^{2-}} ions (omitted for clarity) occupy the vertices of the transition-metal polyhedra. \textbf{b} The magnetic sublattice of octahedrally coordinated \ce{V^{3+}} ($3d^2$, $S=1$) ions forms a pyrochlore network of corner-sharing tetrahedra, forming a geometrically frustrated network of corner-sharing tetrahedra when coupled by antiferromagnetic nearest-neighbour exchange interactions. \textbf{c} Representative synchrotron powder X-ray diffraction profiles of the $511$ and $440$ Bragg reflections (ID22, ESRF) collected between $4\leq T\leq300$~K. Unlike conventional furnace-prepared samples, which exhibit characteristic tetragonal peak splitting below approximately $50$~K \cite{uedaMagneticStructuralTransitions1997}, the Bragg peaks of MASS-produced \ce{ZnV2O4} remain sharp and unsplit down to $T=4$~K, consistent with retention of the average cubic crystallographic ground state. Comparison with the instrumental resolution function (dashed line) shows that the Bragg reflections are close to resolution limited, indicating a high degree of sample crystallinity. \textbf{d} Scanning electron microscopy (SEM) image of MASS-produced \ce{ZnV2O4}, showing micron-scale crystallites with well-faceted morphology.}
    \label{fig:cub}
    \thisfloatpagestyle{empty}
\end{figure}

\noindent We first address whether the altered synthesis pathway changes the crystallographic ground state of \ce{ZnV2O4}. Variable-temperature, high-resolution synchrotron powder X-ray diffraction data provide  evidence of distinct structural behaviour in MASS-produced \ce{ZnV2O4}. Fig.~\ref{fig:cub}c shows two characteristic Bragg reflections of MASS-produced \ce{ZnV2O4} measured on the ID22 X-ray powder diffractometer (ESRF) between $T=4-60$~K (see Methods). Over this temperature range, conventional furnace-prepared polycrystalline samples of \ce{ZnV2O4} typically undergo an orbitally driven cubic-to-tetragonal structural distortion, although the precise transition temperature is itself sample dependent \cite{uedaMagneticStructuralTransitions1997,reehuisCrystallographicMagneticStructure2003,leeOrbitalSpinChains2004,vasilievSpecificHeatMagnetic2006,wangPhaseTransitionCritical2015,browneOrbitalMoleculesVanadium2020,singhaStructuralMagneticStudies2019} and the symmetry of the resulting low-temperature crystal structure remains debated \cite{browneOrbitalMoleculesVanadium2020,pardoHomopolarBondFormation2008,reehuisCrystallographicMagneticStructure2003,uedaMagneticStructuralTransitions1997,zhangLocalOrderFrustration2006,jungInfraredPhononStudy2008}. This transition is readily identified through the characteristic tetragonal splitting of high-resolution Bragg reflections such as the 400 and 511 peaks shown in Fig.~\ref{fig:cub}c. In contrast, all Bragg reflections of MASS-produced \ce{ZnV2O4} remain sharp and undistorted upon cooling to $T=4$~K, establishing that changing the synthesis pathway stabilises the average cubic crystallographic ground state. While previous studies have shown that microwave-assisted synthesis provides rapid access to highly crystalline strongly correlated materials \cite{Ortiz2022}, the present results further demonstrate that the choice of synthesis pathway can also influence the resulting crystallographic ground state. \\

\noindent In order to gain further insight into the structural ground state of MASS-produced \ce{ZnV2O4}, we performed a simultaneous Rietveld refinement of the cubic $Fd\bar{3}m$ spinel model to high-resolution synchrotron powder X-ray (ID22, ESRF) and neutron (HRPD, ISIS) diffraction data collected at $T=10$~K (see Methods). The complementary scattering contrast provided by X-rays and neutrons enables a more robust determination of the average crystal structure than either technique alone, with their distinct scattering contrasts---particularly for vanadium and oxygen---allowing accurate refinement of atomic positions and site occupancies. The simultaneous refinement confirms the cubic symmetry of the sample and yields the structural parameters listed in Table~\ref{tab:crystal}, with the corresponding Rietveld plots shown in Fig.~\ref{fig:Riet} (see Supplementary Information). The model shown in Table~\ref{tab:crystal} allows refinement of the isotropic atomic displacement parameters ($U_{\mathrm{iso}}$) for each ion together with the position of the O$^{2-}$ ion located on the Wyckoff $32e$ $(x,x,x)$ site. In the first instance, this minimal model does not allow for cation site mixing or changes in stoichiometry. It provides an excellent fit to the data over the entire measured temperature range. \\

\noindent To determine whether site mixing or vacancy defects contribute to the distinct structural behaviour of MASS-produced \ce{ZnV2O4}, we next tested several additional structural models, including unconstrained refinement of all site occupancies together with Zn$^{2+}$/V$^{3+}$ site mixing (see Methods and Supplementary Information). No evidence for site mixing was found, with these models failing to improve the fit metrics, as expected from the favourable crystal field stabilisation energy of V$^{3+}$ on the octahedral site of the spinel structure \cite{burnsMineralogicalApplicationsCrystal1993}. Similarly, off-stoichiometric models with varying site occupancies converged to chemically unfeasible solutions. Taken together, these refinements establish that the average crystallographic ground state of MASS-produced \ce{ZnV2O4} remains cubic while providing no evidence for global symmetry lowering, substantial cation site mixing or vacancy disorder within the sensitivity of the combined high-resolution X-ray and neutron diffraction data. Importantly, however, Bragg diffraction is uniquely determined by the spatially averaged crystallographic structure and therefore cannot exclude the presence of local structural distortions, site disorder or vacancies that average out over long-range length scales. Consequently, while the present measurements provide strong evidence that the average cubic ground state is stabilised by the MASS route, they do not preclude local symmetry breaking or short-range structural correlations. Resolving whether such local disorder exists, and whether it contributes to the distinct magnetic ground state discussed below, will require complementary local structural investigations, for example, X-ray pair-distribution-function analysis over a wider Q-range. \\ 

\noindent The simultaneous refinement at $T=10$~K does, however, identify a minor $9.30(6)$~wt.~\% monoclinic \ce{VO2} impurity phase. This most likely arises from the volatility of the \ce{ZnO} precursor during synthesis, leaving residual \ce{V2O3} that is subsequently oxidised to \ce{VO2}. Importantly, the Bragg reflections associated with the \ce{VO2} phase are resolution limited, indicating that it is present as a well-crystallised bulk secondary phase rather than nanoscale domains incorporated within the \ce{ZnV2O4} crystallites. Monoclinic \ce{VO2} undergoes a metal--insulator transition (MIT) at $T=300$~K between a paramagnetic metallic phase above the MIT and a non-magnetic insulating phase below it \cite{okaPhaseTransitionV4+V41993}, ruling out the possibility that this impurity is responsible for the distinct low-temperature magnetic behaviour discussed below. Consistent with this interpretation, optimisation of the MASS synthesis procedure subsequently reduced the \ce{VO2} impurity phase to below the limit of detection by laboratory powder X-ray diffraction without significantly altering either the magnetic susceptibility or structural properties of MASS-produced \ce{ZnV2O4} (Figs.~\ref{fig:synth_dep_2} and \ref{fig:synth_dep_3}, Supplementary Information). Together, the reproducibility of these measurements across independently prepared MASS samples indicate that the distinct structural and magnetic behaviour reported here is reproducible and intrinsic to the \ce{ZnV2O4} phase. \\

\noindent The present results also place the earlier single-crystal study of Ebbinghaus \textit{et al.} \cite{ebbinghausCrystalStructureMagnetic2004} into a broader synthesis-dependent context. Site occupancy refinement of single-crystal X-ray diffraction data collected at $T=100$~K likewise showed no discernible Zn$^{2+}$/V$^{3+}$ site mixing. Furthermore, while no diffraction measurements were reported below $100$~K, the absence of the characteristic anomalies in the magnetic susceptibility led the authors to infer that the low-temperature crystal structure retained cubic symmetry down to $T=2$~K \cite{ebbinghausCrystalStructureMagnetic2004}. Alongside the present study, these observations indicate that suppression of the crystallographically resolved cubic-to-tetragonal transition is not unique to the MASS route, but can arise from distinct synthesis pathways. However, in contrast to previous studies, the present work provides a comprehensive low-temperature characterisation of such an alternative synthesis route, establishing the persistence of the average cubic crystallographic ground state to $T=4$~K and relating this directly to the resulting magnetic behaviour. \\

\noindent These observations suggest that the synthesis pathway influences the structural ground state through more subtle aspects of the microstructure than can be explained by cation disorder or off-stoichiometry alone. Comparison of the FWHM of the measured Bragg peaks of MASS-produced \ce{ZnV2O4} with the instrumental resolution of ID22 (Fig.~\ref{fig:cub}c) reveals near resolution-limited peak profiles, indicating a very high degree of crystallinity achieved \textit{via} the MASS route. Scanning electron microscopy (SEM, Fig.~\ref{fig:cub}d) shows that MASS-produced \ce{ZnV2O4} consists of large, well-separated and highly faceted crystallites. This contrasts with typical furnace-prepared samples, which comprise smaller, highly agglomerated crystallites, highlighting clear differences in particle size and morphology between the two synthesis routes (Fig.~\ref{fig:SEM} and \ref{fig:PSA}, Supplementary Information). Indeed, crystallite size is known to influence the structural behaviour of other spinel systems \cite{prabhuGrainSizeEffect2011,singhSynthesisStructureMorphology2016}. In addition to crystallite size and morphology, an isotropic microstrain broadening of $\epsilon=1.97(3)\times10^{-4}$ was required to reproduce the observed Bragg peak profiles. Taken together, these observations suggest that residual strain and microstructural differences introduced by the synthesis pathway may contribute to stabilising the average cubic ground state. Future high-resolution transmission electron microscopy studies could provide direct insight into the nature and distribution of these strain fields and other microstructural defects \cite{hytchObservingMeasuringStrain2014,linAnalyticalTransmissionElectron2021,wangAtomicScaleStrainField2025}, enabling a direct comparison with conventional furnace-prepared tetragonal samples.

\begin{table}[hbt!]
    \caption{\textbf{Crystal data and structural refinement parameters for MASS-produced \ce{ZnV2O4}.} Structural parameters obtained from simultaneous Rietveld refinement of the cubic crystallographic model against high-resolution neutron powder diffraction (HRPD, ISIS) and synchrotron powder X-ray diffraction (ID22, ESRF) data collected at $T=10$~K. Corresponding Rietveld refinement plots are shown in Fig.~\ref{fig:Riet} (Supplementary Information).}
\centering
\begin{tabular}{lc}
    \\
    \toprule
     & MASS-produced \ce{ZnV2O4}\\
    \midrule
    Crystal system & Cubic \\
    Space group & $Fd \bar{3} m$ \\
    $a$ ($\mathrm{\text{\AA}}$) & 8.39638(1) \\
    $V$ ($\mathrm{\text{\AA}}^3$) &  591.938(1) \\ 
    Crystallite size, $D$ (nm) & 340(2) \\
    Microstrain, $\epsilon$ ($\times10^{-4}$) & 1.97(3) \\
    Zn $U_{\mathrm{iso}}$ ($\mathrm{\text{\AA}}^2$) & 0.00268(6) \\
    V $U_{\mathrm{iso}}$ ($\mathrm{\text{\AA}}^2$) & 0.00171(5) \\
    O $U_{\mathrm{iso}}$ ($\mathrm{\text{\AA}}^2$) & 0.00585(16) \\
    $x$(O) & 0.23927(6) \\
    $T$ (K) & 10 \\
    GOF & 2.558 \\
    $R_{wp} (\%)$ & 13.139 \\
    \bottomrule
\end{tabular}   
\label{tab:crystal}
\end{table}

\FloatBarrier

\section*{Magnetic susceptibility of MASS-produced \ce{ZnV2O4}}

\begin{figure}
    \centering
    \includegraphics[width=0.95\textwidth]{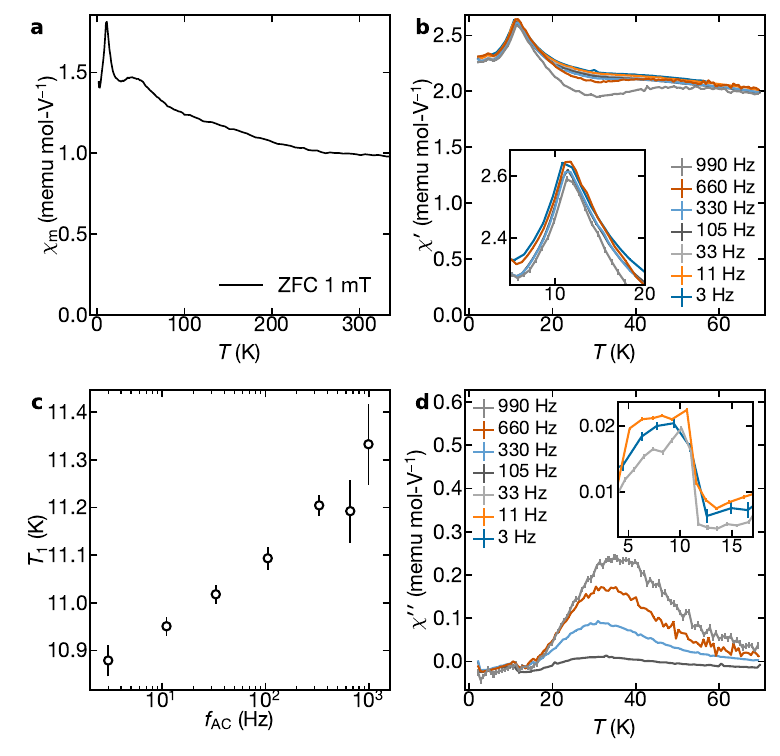}
    \caption{\textbf{Temperature evolution and frequency dependence of the magnetic susceptibility of MASS-produced \ce{ZnV2O4}.} \textbf{a} Temperature dependence of the molar magnetic susceptibility, $\chi_{\mathrm{m}}(T)$, measured in a $1$~mT DC magnetic field under zero-field-cooled conditions, revealing two characteristic magnetic regimes: a sharp cusp at $T_1\approx11$~K and a broad maximum centred around $T_2\approx40$~K. To probe the magnetic dynamics spanning these temperature regimes, AC susceptibility measurements were performed over the frequency range $f_{\mathrm{AC}}=3$--$990$~Hz. \textbf{b} Real component of the AC susceptibility, $\chi^\prime(T)$, showing frequency dependence both around $T_2$ and near $T_1$ (inset). \textbf{c} Frequency dependence of the temperature of the low-temperature cusp, $T_1$, determined from fits to $\chi^\prime(T)$ over the range $10<T<14$~K. \textbf{d} Imaginary component of the AC susceptibility, $\chi^{\prime\prime}(T)$, also showing frequency-dependent behaviour around $T_2$ together with a low-temperature feature near $T_1$ (inset).}
    \label{fig:ACDC_chi}
\end{figure}

\noindent Having established that the MASS pathway stabilises the average cubic crystallographic ground state of \ce{ZnV2O4}, we next examine how this altered structural state influences its magnetic properties. In particular, the persistence of the cubic spinel structure provides an opportunity to investigate the low-temperature magnetic behaviour of a highly frustrated $S=1$ cubic pyrochlore antiferromagnet \cite{gaoSpin1PyrochloreAntiferromagnets2020, iqbalQuantumClassicalPhases2019, pohleAbundanceSpinLiquids2025}. Fig.~\ref{fig:ACDC_chi}a shows the temperature dependence of the molar magnetic susceptibility ($\chi_{\mathrm{m}}(T)$, see Methods) of MASS-produced \ce{ZnV2O4}. $\chi_{\mathrm{m}}(T)$ is commonly modelled using the Curie-Weiss law, \\
\[\chi_{\mathrm{m}}(T)=\frac{C}{T-\theta},\] \\
where $C$ is the Curie constant, proportional to the effective magnetic moment squared, $\mu^2_{\mathrm{eff}}$, and $\theta$ is the Weiss temperature, which reflects the dominant magnetic exchange interactions within a mean-field description. In \ce{ZnV2O4}, however, the magnitude of the antiferromagnetic interactions is sufficiently large that the experimentally accessible temperature range often lies below $|\theta|$ and therefore within the correlated regime, making conventional Curie-Weiss analysis difficult to apply reliably \cite{uedaMagneticStructuralTransitions1997,reehuisCrystallographicMagneticStructure2003,leeOrbitalSpinChains2004,vasilievSpecificHeatMagnetic2006,wangPhaseTransitionCritical2015,browneOrbitalMoleculesVanadium2020,singhaStructuralMagneticStudies2019}. Possibly, the most accurate previous estimate of $\theta$ was obtained by Ueda \textit{et al.}, who fitted susceptibility data collected over the range $400 \leq T \leq 600$~K to obtain $\theta=-420$~K and $\mu_{\mathrm{eff}}=2.73~\mu_{\mathrm{B}}$ per V$^{3+}$. These values are consistent with the high-temperature series-expansion analysis of Nishiguchi \textit{et al.} \cite{nishiguchiPseudotetramerGeometricallyFrustrated2002}, which yielded $\mu_{\mathrm{eff}}=2.77(9)~\mu_{\mathrm{B}}$ per V$^{3+}$ and a nearest-neighbour exchange constant, $J=-92(7)$~K, corresponding to $\theta\approx-370$~K. We have chosen, therefore, to model $\chi_{\mathrm{m}}(T)$ of MASS-produced \ce{ZnV2O4} using a Husimi ansatz (see Methods) \cite{pohleCurielawCrossoverSpin2023}, which has recently been shown to describe the crossover from Curie--Weiss behaviour to the correlated regime in frustrated magnetic materials (see Supplementary Information). This analysis yields $\theta=-404(21)$~K and $\mu_{\mathrm{eff}}=2.76(6)~\mu_{\mathrm{B}}$ per V$^{3+}$ (see Fig.~\ref{fig:DC_chiT} in Supplementary Information), consistent with the previous results of Ueda and Nishiguchi \textit{et al.} \cite{uedaMagneticStructuralTransitions1997,nishiguchiPseudotetramerGeometricallyFrustrated2002} and the expected spin-only moment of $\mu_{\mathrm{eff}}=2.83~\mu_{\mathrm{B}}$ for $S=1$. Finally, no contribution from the minor \ce{VO2} impurity phase is observed in the $\chi_{\mathrm{m}}(T)$ data in Fig.~\ref{fig:ACDC_chi}a, consistent with this secondary phase remaining in its insulating, non-magnetic ground state below the metal--insulator transition \cite{okaPhaseTransitionV4+V41993}. This further supports the conclusion that the impurity phase does not contribute to the low-temperature magnetic behaviour of \ce{ZnV2O4}. \\

\noindent At low temperatures, $\chi_{\mathrm{m}}(T)$ of MASS-produced \ce{ZnV2O4} behaves very differently to that reported for conventional furnace-prepared samples. Fig.~\ref{fig:ACDC_chi}a reveals two characteristic magnetic temperature scales: a sharp cusp at $T_1\approx11$~K and a broader maximum centred around $T_2\approx40$~K, whereas conventional furnace-prepared polycrystalline samples typically exhibit a sharp discontinuity at a sample-dependent temperature in the range $38<T<57$~K (see Table~\ref{tab:ZVO_Ts} in Supplementary Information), corresponding to the cubic-to-tetragonal structural transition at $T_{\mathrm{S}}$, followed by a broader feature centred between $31<T<45$~K \cite{uedaMagneticStructuralTransitions1997,reehuisCrystallographicMagneticStructure2003,leeOrbitalSpinChains2004,vasilievSpecificHeatMagnetic2006,wangPhaseTransitionCritical2015,browneOrbitalMoleculesVanadium2020,singhaStructuralMagneticStudies2019}, which has commonly been associated with the onset of long-range antiferromagnetic order. The absence of the structural transition in MASS-produced \ce{ZnV2O4} gives rise to a qualitatively different susceptibility response, characterised by two distinct magnetic temperature scales rather than the coupled structural and magnetic anomalies observed in conventional samples. \\

\noindent A sharp cusp in $\chi_{\mathrm{m}}(T)$ at $T_1\approx11$~K is also observed in the magnetic susceptibility of the single-crystal \ce{ZnV2O4} reported by Ebbinghaus \textit{et al.} \cite{ebbinghausCrystalStructureMagnetic2004}, where it was proposed that the persistence of the highly frustrated cubic spinel structure gives rise to spin-glass behaviour. Consistent with this interpretation, the magnetic susceptibility of MASS-produced \ce{ZnV2O4} measured in an AC magnetic field (Fig.~\ref{fig:ACDC_chi}b) shows that the position of $T_1$ depends on the frequency of the driving field, $f_\mathrm{AC}$ (Fig.~\ref{fig:ACDC_chi}c), consistent with a spin-freezing transition. However, this spin-glass-like behaviour does not fully account for the low-temperature magnetic susceptibility, given the presence of the broad feature centred at $T_2\approx40$~K. Furthermore, the AC susceptibility reveals pronounced dynamical behaviour around this temperature over the frequency range $150<f_\mathrm{AC}<990$~Hz (Fig.~\ref{fig:ACDC_chi}d), indicating that the magnetic response of MASS-produced \ce{ZnV2O4} cannot be understood solely in terms of a conventional spin-glass transition. The frequency dependence of both $T_1$ and $T_2$ can be described by an activated Arrhenius form, with the corresponding analysis presented in Sec.~\ref{sec:AC} of the Supplementary Information. To further explore these unusual magnetic dynamics, we next turn to inelastic neutron scattering, which provides direct insight into the magnetic correlations and excitation spectrum of MASS-produced \ce{ZnV2O4}. \\

\FloatBarrier

\section*{Magnetic excitation spectrum of MASS-produced \ce{ZnV2O4}}

\begin{figure}[hbt!]
    \centering
    \includegraphics[width=8.9cm]{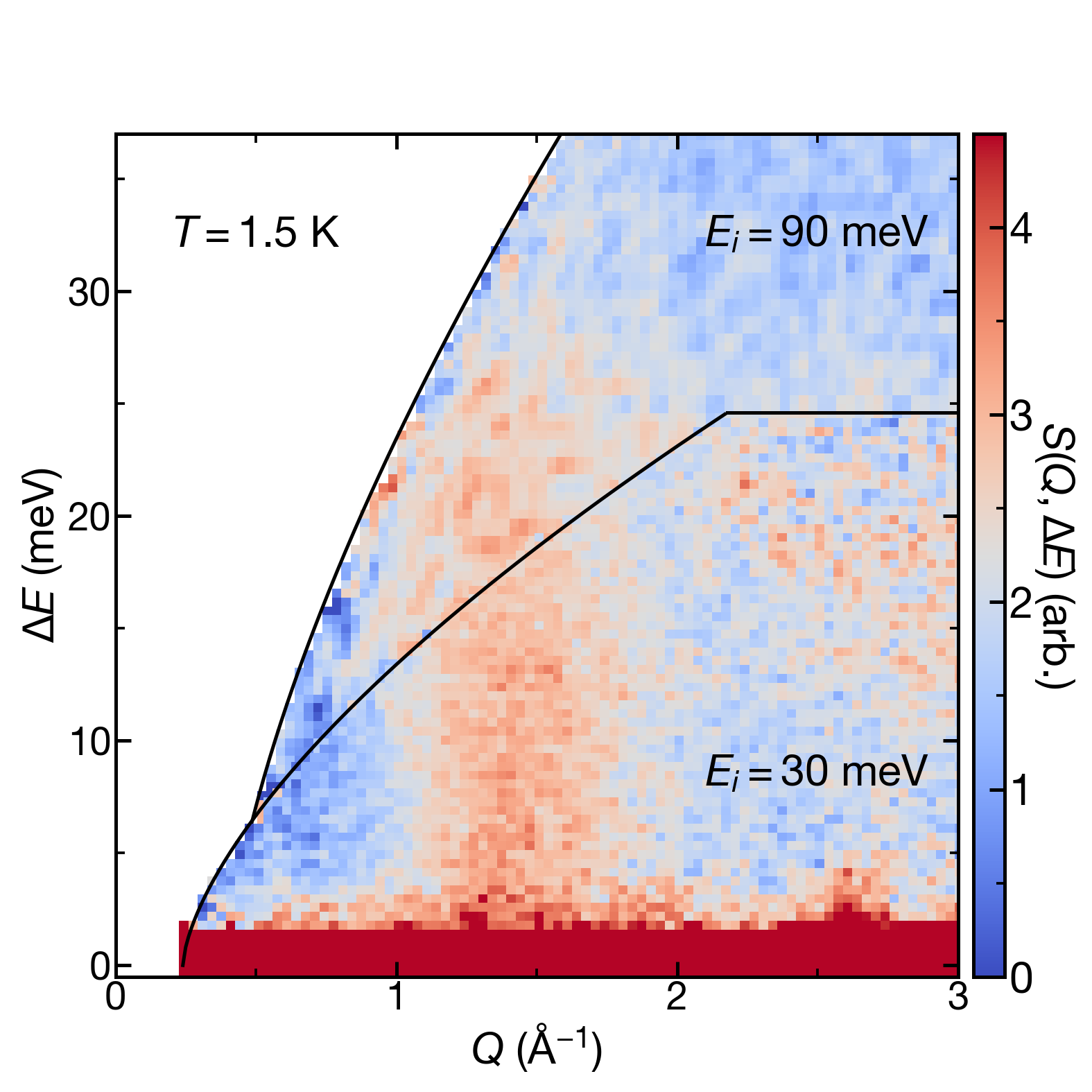}
    \caption{\textbf{Magnetic excitation spectrum of MASS-produced \ce{ZnV2O4}.} Dynamical structure factor, $S(Q,\Delta E)$, of MASS-produced \ce{ZnV2O4} measured by inelastic neutron scattering on the PANTHER spectrometer (ILL) at $T=1.5$~K using incident neutron energies of $E_i=30$ and $90$~meV. The data reveal a broad column of diffuse magnetic scattering centred at $Q_\mathrm{AF}\approx1.4$~\text{\AA}$^{-1}$ that extends to at least $\Delta E\approx30$~meV. This indicates the presence of persistent dynamic spin fluctuations associated with strong short-range antiferromagnetic correlations over the nearest-neighbour V--V length scale. The feature close to the elastic line at $Q\approx2.6$~\text{\AA}$^{-1}$ is a structural Bragg peak, while the broad scattering centred at $\Delta E\approx20$~meV at higher $Q$ arises from incoherent phonon scattering.} 
    \label{fig:INS}
\end{figure}

\noindent To investigate the magnetic correlations and spin dynamics of MASS-produced \ce{ZnV2O4}, we performed a comprehensive inelastic neutron scattering (INS) study over a wide dynamic range and multiple energy resolutions (see Methods). Figure~\ref{fig:INS} shows the momentum- and energy-transfer dependence of the dynamical structure factor, $S(Q,\Delta E)$, measured at $T=1.5$~K on the PANTHER time-of-flight spectrometer (ILL) using incident neutron energies of $E_i=30$ and $90$~meV (see Methods). In a magnet exhibiting conventional long-range magnetic order, INS measurements typically reveal dispersive spin-wave excitations reflecting coherent collective spin precession \cite{Toth_2015}. Instead, MASS-produced \ce{ZnV2O4} exhibits a broad column of diffuse inelastic magnetic scattering centred at $Q_\mathrm{AF}\approx1.4$~\text{\AA}$^{-1}$ that extends continuously to at least $\Delta E=30$~meV (Fig.~\ref{fig:INS}). This diffuse inelastic response indicates that the magnetic moments remain dynamic while exhibiting strong short-range antiferromagnetic correlations rather than developing coherent long-range magnetic order. The scattering is centred at a momentum transfer corresponding approximately to first near-neighbour antiferromagnetic correlations over a real-space distance of $r_{\mathrm{AF}}\simeq4.5/Q_\mathrm{AF}\approx3.2$~\text{\AA}, close to the nearest-neighbour V--V separation of $2.97$~\AA~at $T=1.5$~K (see Fig.~\ref{fig:chi2_lor} in the Supplementary Information). Although the full bandwidth of these magnetic excitations cannot be determined because the scattering extends beyond the kinematic range accessible in the present measurements, the observed response extends to at least $30$~meV, corresponding to an energy scale of approximately $350$~K. \\

\noindent While the high incident-energy measurements on PANTHER are essential for characterising the broad magnetic excitation spectrum extending to at least $30$~meV, they do not provide sufficient energy resolution to resolve the slow spin dynamics associated with the frequency-dependent anomaly observed in the magnetic susceptibility at $T_1\approx11$~K. We therefore complemented the PANTHER measurements with low-energy INS measurements performed on the LET time-of-flight spectrometer at the ISIS Neutron and Muon Source (see Methods). Together, these measurements provide access to magnetic excitations over a combined energy-transfer range of $0.4 \leq |\Delta E| \leq 30$~meV at $Q_\mathrm{AF}$ with an energy resolution of $120~\mathrm{\mu eV}$, enabling both the fast and slow magnetic dynamics of MASS-produced \ce{ZnV2O4} to be characterised within a single, self-consistent framework. \\

\begin{figure}[hbt!]
    \centering
    \includegraphics[width=0.95\textwidth]{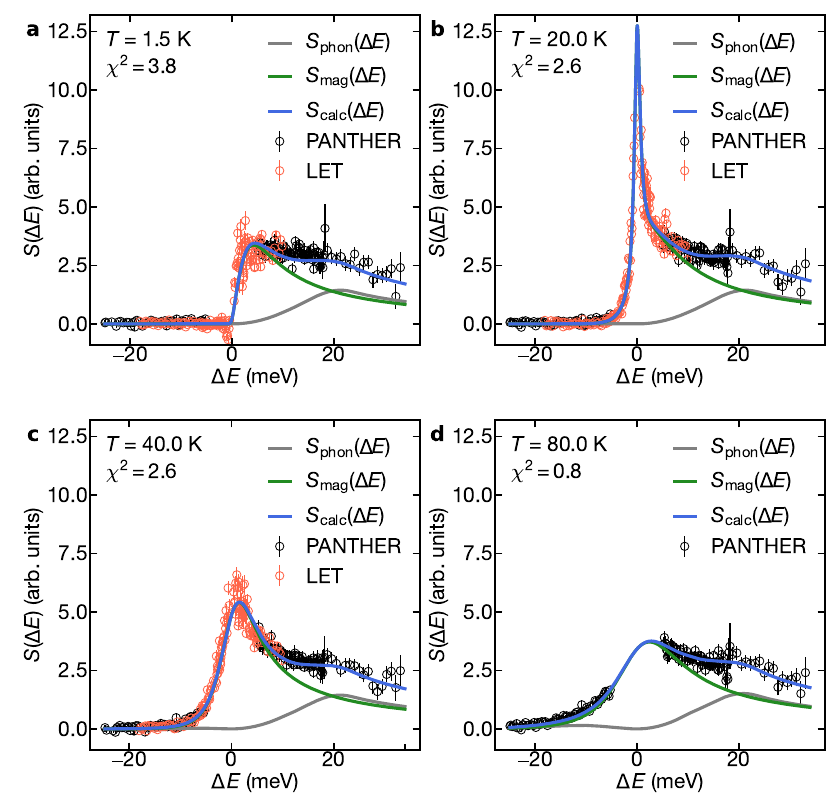}
    \caption{\textbf{Modelling fast and slow magnetic fluctuations in MASS-produced \ce{ZnV2O4}.} Combined (see Methods) $S(\Delta E)$ data at $Q_\mathrm{AF}$, integrated over $1.35 \leq Q \leq 1.45$~\AA$^{-1}$, obtained from INS measurements on the PANTHER (ILL) and LET (ISIS) spectrometers at \textbf{a} $T=1.5$~K, \textbf{b} $20$~K, \textbf{c} $40$~K and \textbf{d} $80$~K. The solid lines show simultaneous fits to Eq.~\ref{eq:Scalc} (see Methods), excluding the elastic contribution, $I_\mathrm{el}(Q)$. The measurement temperature and reduced-$\chi^2$ goodness-of-fit are indicated in the upper-left corner of each panel. The magnetic response comprises a broad inelastic component, $\chi''_2$, present at all measured temperatures, together with an additional narrow inelastic component, $\chi''_1$, that is observed only over a limited temperature range close to $T_1$ and is most clearly resolved at $20$~K (\textbf{b}). Data in \textbf{d} are from PANTHER only.}
    \label{fig:INS_LP}
\end{figure}

\noindent Inelastic magnetic neutron scattering provides a direct measurement of the imaginary, or dissipative, component of the dynamic susceptibility, $\chi''(Q,\Delta E)$, according to the fluctuation--dissipation theorem of Kubo~\cite{kuboFluctuationdissipationTheorem1966},
\[
S_{\mathrm{mag}}(Q,\Delta E) \propto [1+n(\Delta E)]\chi''(Q,\Delta E),
\]
where $n(\Delta E)$ is the Bose population factor. To isolate the magnetic contribution to the INS spectra of MASS-produced \ce{ZnV2O4}, the structural phonon contribution must first be accounted for. We therefore measured the non-magnetic, isostructural analogue \ce{ZnGa2O4}, from which a phonon model for \ce{ZnV2O4} was derived (see Methods). Using a self-consistent modelling approach across the INS datasets (see Methods and Supplementary Information Section~\ref{sec:INS_phon}), we find that the resulting $S(Q,\Delta E)$ can be described by the expression given in Eq.~\ref{eq:Scalc}. The magnetic response is represented by two overdamped harmonic-oscillator components, $\chi_1''(Q,\Delta E)$ and $\chi_2''(Q,\Delta E)$, which capture fluctuations on distinct characteristic energy scales. We find no evidence for a magnetic excitation gap at any measured temperature. Fitting this model to data collected over $T=1.5-300$~K (Fig.~\ref{fig:INS_LP}) yields the temperature- and $Q$-dependent static susceptibilities, $\chi_{1,2}(Q,T)$, and characteristic spin-relaxation rates, $\Gamma_{1,2}(Q,T)$, of the two components. \\

\begin{figure}[hbt!]
    \centering
    \includegraphics[width=0.9\textwidth]{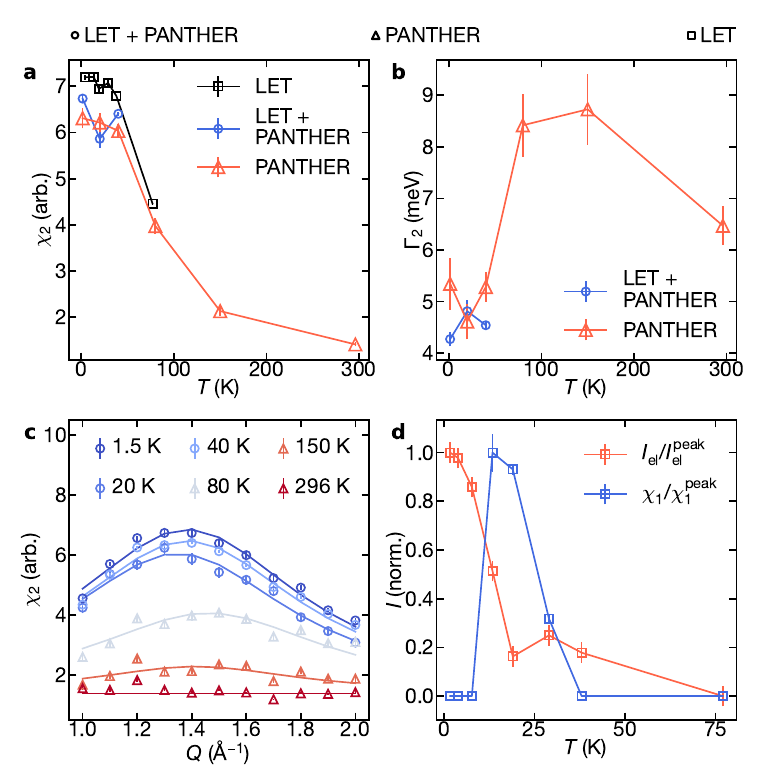}
    \caption{\textbf{Temperature evolution of the magnetic dynamics of MASS-produced \ce{ZnV2O4}.} Fitting the magnetic component of the dynamical structure factor, $S_{\mathrm{mag}}(Q,\Delta E)$, extracted from INS measurements on PANTHER (ILL) and LET (ISIS) at $Q_\mathrm{AF}$ to the model given in Eq.~\ref{eq:Scalc} yields \textbf{a} the static susceptibility, $\chi_2(Q_\mathrm{AF},T)$, and \textbf{b} the spin relaxation rate, $\Gamma_2(Q_\mathrm{AF},T)$, associated with the broad magnetic response. Panels \textbf{a} and \textbf{b} compare values obtained from independent fits to the PANTHER data (triangles), LET data (squares) and simultaneous fits to the combined datasets (circles), demonstrating excellent agreement between the independent analyses. Both $\chi_2(Q_\mathrm{AF},T)$ and $\Gamma_2(Q_\mathrm{AF},T)$ become independent of temperature below $T\approx40$~K. \textbf{c} The $Q$-dependence of $\chi_2(Q,T)$ is well described by a Lorentzian lineshape (solid lines), from which a short spin correlation length, $\xi$, is obtained that is approximately temperature independent and comparable to the nearest-neighbour V--V separation. In contrast, \textbf{d} the static susceptibility associated with the low-energy magnetic response observed on LET, $\chi_1(Q_\mathrm{AF},T)$, is present only between $T=13$ and $30$~K. Its disappearance below $T\approx13$~K coincides with an increase in the elastic magnetic scattering intensity, $I_\mathrm{el}$, indicating that the slowest magnetic fluctuations become static on the timescale of the neutron scattering experiment.}
    \label{fig:INS_metrics}
    \thisfloatpagestyle{empty}
\end{figure}

\noindent Figs.~\ref{fig:INS_metrics}a and b show the temperature dependence of the fitted static susceptibility, $\chi_2(T)$, and relaxation rate, $\Gamma_2(T)$, associated with the $\chi_2''(Q,\Delta E)$ response at $Q_\mathrm{AF}$. The values shown are derived independently from the modelling of the PANTHER, LET, and combined PANTHER + LET datasets, and their close agreement demonstrates the robustness of the extracted magnetic response across multiple instrumental datasets and energy ranges. Both $\chi_2(Q_\mathrm{AF})$ and $\Gamma_2(Q_\mathrm{AF})$ become independent of temperature below $T\approx40$~K, coinciding with the broad maximum observed in the AC and DC magnetic susceptibilities of MASS-produced \ce{ZnV2O4} (Figs.~\ref{fig:ACDC_chi}a,b,d). The $Q$-dependent static susceptibility, $\chi_2(Q)$, shown in Fig.~\ref{fig:INS_metrics}c is likewise independent of temperature for all measured $Q$ below $T\approx40$~K, while strong nearest-neighbour magnetic correlations persist to at least $T=150$~K. The corresponding spin correlation length, $\xi(T)$, extracted by fitting $\chi_2(Q,T)$ to a Lorentzian function (Fig.~\ref{fig:INS_metrics}c and Supplementary Information Fig.~\ref{fig:chi2_lor}), is also remarkably temperature independent between $1.5\leq T\leq150$~K, with an average value of $\xi=2.0(4)$~\AA, shorter than the nearest-neighbour V--V separation of $2.97$~\AA. Together, these observations indicate that MASS-produced \ce{ZnV2O4} remains dominated by dynamic, short-range antiferromagnetic correlations over a wide temperature range, despite the strong magnetic interactions implied by the large negative Weiss temperature. Furthermore, the low-temperature limit of $\Gamma_2(Q_\mathrm{AF})\approx5$~meV corresponds to a characteristic energy scale of approximately $58$~K, comparable to $T_2\approx40$~K. Inversion of $\Gamma_2(Q_\mathrm{AF})$ (see Methods) gives a spin-fluctuation lifetime of $\tau\approx0.16$~ps. By comparison, a purely thermal spin fluctuation at $T=1.5$~K would be expected to have a much longer lifetime of approximately $3.3$~ps. These observations indicate that the low-temperature magnetic response is dominated by persistent, temperature-independent spin fluctuations rather than thermally activated spin dynamics. \\

\noindent In contrast to the $\chi_2''(Q,\Delta E)$ response, the low-energy component, $\chi_1(Q_\mathrm{AF})$, is observed only in the high-resolution LET measurements over a narrow temperature interval between $T=13$ and $30$~K (Fig.~\ref{fig:INS_metrics}d). Below $T\approx13$~K, this inelastic response disappears, coincident with a sharp increase in the elastic magnetic intensity, $I_\mathrm{el}$. This transfer of spectral weight from the inelastic to the elastic channel indicates that the slowest magnetic fluctuations become static on the timescale of the neutron scattering experiment. The onset of this behaviour occurs at a temperature consistent with the frequency-dependent cusp observed in the AC magnetic susceptibility at $T_1\approx11$~K (Figs.~\ref{fig:ACDC_chi}b,c). \\

\noindent Taken together, the INS measurements reveal two distinct dynamical components with markedly different temperature dependences. The higher-energy response remains dynamic and essentially temperature independent below $T_2\approx40$~K, whereas the lower-energy response disappears below $T_1\approx11$~K as spectral weight is transferred into the elastic channel. Viewed in the context of the predicted phase diagram of the $S=1$ pyrochlore antiferromagnet~\cite{pohleAbundanceSpinLiquids2025} and by comparison with other frustrated magnetic systems discussed below, one possible interpretation is that the higher-energy response is associated with quantum spin-liquid-like fluctuations, while the lower-energy response reflects the freezing of the slowest magnetic fluctuations. Immediately above the freezing transition ($T=20$~K), the ratio of $\chi_2(Q_\mathrm{AF})$ to $\chi_1(Q_\mathrm{AF})$ indicates that approximately $63$~\% of the magnetic scattering is associated with the higher-energy dynamical response. \\

\noindent Comparing the INS spectrum of MASS-produced \ce{ZnV2O4} with the only previous INS study of a furnace-prepared sample further highlights the distinct low-temperature magnetic behaviour associated with the synthesis-pathway-stabilised cubic ground state \cite{leeOrbitalSpinChains2004}. At $T=100$~K, the INS spectrum of furnace-prepared \ce{ZnV2O4} likewise exhibits a column of diffuse magnetic scattering centred at $Q\approx1.35$~\text{\AA}$^{-1}$ that extends to the maximum accessible energy transfer of the reported measurements ($\Delta E\approx17$~meV) \cite{leeOrbitalSpinChains2004}, similar to that observed for MASS-produced \ce{ZnV2O4} at $T=1.5$~K (Fig.~\ref{fig:INS}). Below the cubic-to-tetragonal structural transition at $T<50$~K, however, the energy-integrated scattering develops a highly asymmetric $Q$-dependence that was well described by a one-dimensional spin-chain model, consistent with orbital ordering in the tetragonal phase \cite{leeOrbitalSpinChains2004}. At $T<40$~K, the magnetic spectral weight transfers into the elastic channel, opening an excitation gap below $\Delta E<9$~meV \cite{leeOrbitalSpinChains2004}. This contrasts sharply with the broad, gapless magnetic excitation spectrum and persistent dynamic spin fluctuations observed in MASS-produced \ce{ZnV2O4} at $T=1.5$~K. Strikingly, we observe no evidence of the characteristic asymmetric scattering in the measured $\chi_2(Q)$ of MASS-produced \ce{ZnV2O4} (Fig.~\ref{fig:INS_metrics}c), inconsistent with the one-dimensional magnetic correlations reported for the tetragonal phase. Importantly, this absence of asymmetry cannot be attributed to powder averaging: powder-averaged magnetic neutron scattering from a one-dimensional chain, for which the scattering forms continuous planes of intensity in reciprocal space perpendicular to the chain direction, produces a characteristically asymmetric lineshape as a function of $|Q|$, as observed previously \cite{leeOrbitalSpinChains2004,Xu96}. \\

\noindent The magnetic phenomenology of MASS-produced \ce{ZnV2O4} is similar to that reported for the $S=1$ pyrochlore antiferromagnet \ce{NaCaNi2F7} \cite{krizanNaCaNi2F7FrustratedHightemperature2015,plumbContinuumQuantumFluctuations2019}. \ce{NaCaNi2F7} adopts the cubic $Fd\bar{3}m$ pyrochlore structure, in which octahedrally coordinated \ce{Ni^{2+}} ($3d^8$, $S=1$) ions occupy a lattice of corner-sharing tetrahedra, while Na$^+$ and Ca$^{2+}$ are randomly distributed over the non-magnetic $A$-site. In this system, temperature-independent spin fluctuations emerge below approximately $18$~K, followed by the freezing of the slowest spin dynamics below approximately $4$~K \cite{plumbContinuumQuantumFluctuations2019}. This parallels the phenomenology observed in MASS-produced \ce{ZnV2O4}, although the characteristic temperatures are higher, with temperature-independent spin fluctuations emerging below $T_2\approx40$~K and the slowest fluctuations freezing below $T_1\approx11$~K. The higher characteristic temperatures in MASS-produced \ce{ZnV2O4} are consistent with its substantially stronger magnetic exchange interactions and are accompanied by a broader magnetic excitation spectrum extending to at least $30$~meV, compared with approximately $12.5$~meV in \ce{NaCaNi2F7}. Taken together, these similarities are consistent with interpreting the higher-energy response in MASS-produced \ce{ZnV2O4} as arising from quantum spin-liquid-like fluctuations, while the lower-energy response reflects the freezing of the slowest spin dynamics. At the same time, the differences between the two systems may provide important insight into the role of structural disorder. In \ce{NaCaNi2F7}, the partial freezing of the lowest-energy spin fluctuations has been attributed to Na$^+$/Ca$^{2+}$ site disorder \cite{plumbContinuumQuantumFluctuations2019}. \\

\noindent The observation of similar phenomenology in MASS-produced \ce{ZnV2O4}, despite the absence of any evidence for Zn$^{2+}$/V$^{3+}$ site disorder in the present diffraction measurements, suggests that an alternative mechanism may be responsible for the partial freezing we observe. Indeed, recent work by Syzranov and Ramirez~\cite{syzranovEminuscentPhaseFrustrated2022} highlights the ubiquity of spin-freezing transitions in apparently disorder-free geometrically frustrated materials. They refer to this frozen state as an ``eminuscent'' phase, appearing below a characteristic temperature that only decreases with increasing vacancy and substitutional disorder. They speculate that other sources of disorder, such as random strain, as observed in the $S=1$ pyrochlore antiferromagnet \ce{Y2Mo2O7}~\cite{greedanLocalAverageStructures2009,thygesenOrbitalDimerModel2017}, could give rise to this behaviour. In \ce{Y2Mo2O7}, the orbital degeneracy of the octahedrally coordinated Mo$^{4+}$ ($4d^2$) ions resembles that of the octahedrally coordinated V$^{3+}$ ($3d^2$) ions in \ce{ZnV2O4}. Yet, like MASS-produced \ce{ZnV2O4}, the average crystallographic ground state of \ce{Y2Mo2O7} remains cubic \cite{reimersCrystalStructureSpinglass1988,greedanLocalAverageStructures2009}. Total-scattering measurements proved crucial in resolving this apparent paradox, revealing through pair-distribution-function analysis that the local structure deviates from the average $Fd\bar{3}m$ symmetry owing to cooperative two-in--two-out displacements of the Mo$^{4+}$ ions, giving rise to a static orbital-ice state \cite{thygesenOrbitalDimerModel2017}. Future neutron and X-ray total-scattering studies of MASS-produced \ce{ZnV2O4} could therefore directly test whether local orbital symmetry breaking similarly persists despite the average cubic crystallographic ground state established here.

\section*{Conclusions and outlook}

\noindent This work demonstrates that synthesis pathway can act as a powerful design parameter for selecting emergent quantum states in strongly correlated materials. Through the development of a microwave-assisted solid-state synthesis (MASS) route for \ce{ZnV2O4}, we have stabilised a cubic crystallographic ground state that gives rise to markedly different magnetic behaviour from that of conventional furnace-prepared samples. Our combined diffraction, magnetometry and inelastic neutron scattering measurements reveal persistent, temperature-independent spin fluctuations below $T_2\approx40$~K together with the freezing of the slowest spin dynamics below $T_1\approx11$~K. Taken together, these observations establish a fundamentally different magnetic ground state in MASS-produced \ce{ZnV2O4}, in which persistent spin-liquid-like dynamics coexist with partial spin freezing. The persistence of the average cubic crystallographic ground state simultaneously preserves orbital degeneracy and suppresses long-range orbital order, raising the prospect of an accompanying orbital-liquid, orbital-glass or orbital-ice state \cite{thygesenOrbitalDimerModel2017}. \\

\noindent The striking contrast between MASS-produced and conventional furnace-prepared \ce{ZnV2O4} therefore demonstrates that changing synthesis pathway alone can select distinct structural and magnetic ground states. The rapid and scalable nature of MASS additionally enables systematic reproducibility studies, synthetic optimisation, sample scale-up and, potentially, \textit{in situ} investigations of solid-state reaction pathways \cite{McFadzean2025}. Given its demonstrated applicability across a wide range of oxides \cite{Ortiz2022}, chalcogenides \cite{Aitken2007} and intermetallics \cite{Lekse2007}, MASS provides a promising route for testing whether similarly synthesis-selected states can be accessed across broader classes of quantum materials. \\

\noindent Recent theoretical work has revealed the $S=1$ pyrochlore antiferromagnet to host a remarkably rich landscape of competing quantum phases \cite{pohleAbundanceSpinLiquids2025}. Determining the precise nature of the emergent magnetic state realised in MASS-produced \ce{ZnV2O4} will require further experimental and theoretical investigation, particularly through single-crystal inelastic neutron scattering and local structural probes capable of resolving orbital disorder beyond the average crystallographic structure. Equally important is understanding why distinct synthesis pathways stabilise different crystallographic and magnetic ground states. The present work identifies crystallite morphology, size and strain as plausible contributors to this synthesis dependence, providing experimental parameters that have been largely overlooked throughout almost three decades of studies of \ce{ZnV2O4}. A careful comparison of the local structure of MASS-produced and furnace-prepared samples using neutron and X-ray total scattering together with high-resolution electron microscopy therefore represents an important next step. More generally, establishing how synthesis pathway governs local structure and disorder, and how these in turn select emergent behaviour, represents an important challenge---and opportunity---for the future design and discovery of quantum materials.

\FloatBarrier

\section*{Methods}

\subsection*{Sample synthesis}

\noindent Polycrystalline samples of \ce{ZnV2O4} were prepared \textit{via} a rapid microwave-assisted solid-state synthesis (MASS) method. Stoichiometric quantities of \ce{ZnO} (99.999\%, Sigma Aldrich) and \ce{V2O3} (99.99\%, Sigma Aldrich) were ground by hand in a mortar and pestle for 10~minutes. The resulting powder was pressed into pellets of $\approx400$~mg and sealed under vacuum in quartz ampoules ($\leq10^{-3}$ mbar). Each ampoule was embedded in an alumina crucible filled with $\approx3$~g of activated charcoal microwave susceptor. The crucible was encased in aluminium silicate ceramic fibreboard for thermal insulation and heated in a Panasonic NN-SN651B $1200$~W microwave oven at $30$~\% power (equivalent to $360$~W) for 12~minutes. After heating, the charcoal temperature was measured to be $\approx1013$~K using a K-type thermocouple. The sample was allowed to cool to room temperature before being recovered from the ampoule and ground by hand into a fine black powder. In total, $\approx2$~g of MASS-produced \ce{ZnV2O4} was synthesised in multiple serial batches of approximately $400$~mg. Batch-to-batch reproducibility was confirmed by laboratory powder X-ray diffraction and magnetic susceptibility measurements. The high crystallinity of the combined multi-batch sample, as demonstrated by the high-resolution synchrotron powder X-ray diffraction data (Fig.~\ref{fig:cub}c), further confirms the reproducibility of the MASS method. All measurements presented in this study were performed on this combined multi-batch sample. \\

\noindent The MASS method was subsequently optimised to minimise the \ce{VO2} impurity phase present in the samples prepared by the procedure described above. Samples were synthesised following the same protocol, but with up to $10$~\% excess ZnO added to the starting mixture to compensate for ZnO volatility during heating, and with the quartz ampoules flushed and purged with Ar or \ce{N2} prior to sealing to minimise oxidation of the reaction precursors and products. Laboratory powder X-ray diffraction confirmed that these optimised synthesis conditions reproducibly yielded phase-pure samples (Fig.~\ref{fig:synth_dep_2}, Supplementary Information). Magnetic susceptibility measurements further confirmed that the optimised samples reproducibly exhibit the same characteristic magnetic behaviour as the samples studied in the main text, namely a sharp cusp in $\chi_{\mathrm{m}}$ at $T\approx11$~K together with the absence of the structural ($T_{\mathrm{S}}$) and magnetic ($T_{\mathrm{N}}$) transitions characteristic of conventional furnace-prepared samples (Fig.~\ref{fig:synth_dep_3}, Supplementary Information). \\

\noindent To estimate the phonon contribution to the INS spectrum of MASS-produced \ce{ZnV2O4}, a polycrystalline sample of the non-magnetic, isostructural analogue \ce{ZnGa2O4} was prepared by a conventional furnace-heated solid-state synthesis. Stoichiometric quantities of \ce{ZnO} (99.9995\%, Alfa Aesar) and \ce{Ga2O3} (99.999\%, Alfa Aesar) were combined and ground by hand in an agate mortar and pestle with $1$~mL of ethanol for 5~minutes. The resulting slurry was dried in air for $3$~hours before the powder was pressed into $\approx500$~mg pellets. The pellets were fired in air at $1473$~K for $24$~hours and allowed to cool to room temperature before being ground by hand into a fine white powder. Batch-to-batch reproducibility was confirmed by laboratory powder X-ray diffraction, yielding a total sample mass of approximately $3.5$~g.

\subsection*{High-resolution powder diffraction and crystal structure refinement}\label{PXRD_method}

\noindent High-resolution synchrotron powder X-ray diffraction (PXRD) data were collected for MASS-produced \ce{ZnV2O4} on the ID22 beamline at the European Synchrotron Radiation Facility (ESRF) \cite{fitchID22HighresolutionPowderdiffraction2023}. The sample was packed into $1$~mm borosilicate glass capillaries and cooled using a helium cryostream, with the capillaries left open to allow gas exchange. Data were collected using an Eiger 2 X \ce{CdTe} 2M-W detector with nine Si(111) analyser crystals. The incident X-ray energy was $35$~keV, corresponding to a calibrated wavelength of $\lambda = 0.354195$~\AA. Data were collected in 2~K temperature intervals between $4$ and $55$~K, 5~K intervals between $60$ and $100$~K, and 50~K intervals between $150$ and $300$~K. A Si standard was measured at $300$~K to determine the instrumental resolution, $\Delta d/d = 1.5\times10^{-4}$, by Rietveld analysis using the TOPAS-Academic software package \cite{coelhoTOPASTOPASAcademicOptimization2018}. \\

\noindent High-resolution time-of-flight powder neutron diffraction (PND) data were collected for MASS-produced \ce{ZnV2O4} on the HRPD diffractometer at the ISIS Neutron and Muon Source \cite{ibbersonDesignPerformanceNew2009}. The sample was measured in a flat-plate vanadium-windowed holder ($18\times23\times2$~mm) mounted in a top-loading closed-cycle refrigerator between $10$ and $120$~K in $10$~K intervals. A Si standard was also measured at $300$~K to determine the instrumental resolution (Bank 1: $\Delta d/d = 5.3\times10^{-4}$, Bank 2: $\Delta d/d = 2.3\times10^{-3}$, Bank 3: $\Delta d/d = 8.0\times10^{-3}$). \\

\noindent The average crystallographic ground-state structure of MASS-produced \ce{ZnV2O4} was determined by simultaneous Rietveld refinement of the cubic $Fd\bar{3}m$ spinel model against the PXRD data collected on ID22 and the PND data collected on HRPD Banks 1, 2 and 3 at $T=10$~K (Fig.~\ref{fig:Riet}, Supplementary Information). The refined parameters for the cubic \ce{ZnV2O4} phase comprised the lattice parameter, oxygen positional parameter, independent isotropic atomic displacement factors for each crystallographic site, isotropic crystallite-size broadening and isotropic microstrain broadening. To assess possible site mixing and deviations from stoichiometry, refinements were also performed allowing the occupancies of the Zn$^{2+}$, V$^{3+}$ and O$^{2-}$ sites to vary, together with Zn$^{2+}$/V$^{3+}$ site mixing, with both fixed and refined O$^{2-}$ occupancies (Tabs.~\ref{tab:riet_1} and \ref{tab:riet_2}, Supplementary Information). All refinements included a secondary monoclinic \ce{VO2} phase (space group $C2/m$), for which the lattice parameters, a single isotropic atomic displacement factor and isotropic crystallite-size broadening were refined. 

\subsection*{Scanning electron microscopy} 

\noindent Scanning electron microscopy (SEM) images were collected for MASS-produced and furnace-prepared \ce{ZnV2O4} samples, the latter synthesised following a conventional literature procedure (\textit{e.g.} Ref.~\cite{reehuisCrystallographicMagneticStructure2003}), using a ZEISS EVO 10 electron microscope. Samples were mounted on carbon tabs and coated with an $8$~nm Au layer using a Quorum Q150RS Plus sputter coater to minimise charging. The microscope was operated at an accelerating voltage of 15~kV and a probe current of $100$~pA. Between 7 and 10 electron micrographs were collected for each sample over a range of magnifications ($500-10,000\times$). The diameters of at least 30 well-isolated particles were measured from each image, giving a total of approximately $200-300$ particle diameters per sample. Image analysis was performed using ImageJ (Version 1.54) to determine the particle-size distributions shown in Supplementary Information Figs.~\ref{fig:SEM} and \ref{fig:PSA}. 

\subsection*{Magnetic susceptibility}

\noindent Magnetic susceptibility measurements of MASS-produced \ce{ZnV2O4} were performed using a Quantum Design MPMS3 SQUID magnetometer. For DC measurements, a $35.54$~mg powder sample was packed into a gelatin capsule and mounted in a standard MPMS3 straw holder. Zero-field-cooled (ZFC) and field-cooled (FC) measurements were collected in an applied magnetic field of $H_\mathrm{DC}=1$~mT over the temperature range $2\leq T\leq350$~K. Prior to each measurement, the residual magnetic field of the superconducting solenoid was nulled using the Quantum Design ultra-low-field option. The reduced susceptibility, $\chi_\mathrm{m}T$, was fitted using a Husimi ansatz \cite{pohleCurielawCrossoverSpin2023} over the temperature range $50\leq T\leq335$~K to determine the Curie and Weiss constants (Supplementary Information, Fig.~\ref{fig:DC_chiT}). \\

\noindent AC susceptibility measurements were performed on a $101.9$~mg powder sample over the temperature range $2\leq T\leq70$~K, spanning both characteristic magnetic regimes ($T_1$ and $T_2$), using an excitation field of $H_\mathrm{AC}=1$~mT at frequencies of $f_\mathrm{AC}=3,\:11,\:33,\:105$ and $330$~Hz. Measurements at $f_\mathrm{AC}=660$ and $990$~Hz were performed using reduced excitation fields of $H_\mathrm{AC}=0.4$ and $0.05$~mT, respectively, owing to the maximum drive amplitude available on the MPMS3 at high frequencies. The AC field amplitude and phase calibration were verified via measurements of superconducting Nb.

\subsection*{Inelastic Neutron Scattering}
\subsubsection*{Data collection}

\noindent Time-of-flight inelastic neutron scattering (INS) spectra of MASS-produced \ce{ZnV2O4} were measured on the PANTHER thermal spectrometer at the Institut Laue-Langevin (ILL), and the LET cold-neutron and MERLIN thermal-neutron spectrometers at the ISIS Neutron and Muon Source. PANTHER was configured to run at incident energies $E_i \in \{30,90\}$~meV with FWHM energy resolutions $\delta E \in \{1.2,5.5\}$~meV. LET measurements were collected simultaneously using repetition-rate multiplication~\cite{russinaFirstImplementationRepetition2009} at $E_i \in \{2.2,3.7,7.5,22.8\}$~meV with FWHM energy resolutions $\delta E \in \{0.06,0.12,0.34,1.41\}$~meV. MERLIN measurements were collected simultaneously using repetition-rate multiplication at $E_i \in \{30,60,180\}$~meV with FWHM energy resolutions $\delta E \in \{1.2,2.8,13.0\}$~meV. \\

\noindent For all three experiments, the $\approx2$~g MASS-produced \ce{ZnV2O4} sample was mounted in an annular cylindrical Al sample holder with a sample thickness of $1$~mm, limiting sample self-attenuation to below 5\% for all incident energies. The temperature was varied between $T=1.5-150$~K on PANTHER ($T=1.5-80$~K on LET) using a standard ``orange'' He-flow cryostat. In addition, spectra at $T=296$~K were collected on PANTHER with the sample loaded directly into the instrument vacuum tank. On MERLIN, the temperature was varied between $T=1.8-300$~K using a standard ``orange'' He-flow cryostat. \\

\noindent INS spectra of non-magnetic \ce{ZnGa2O4} were measured at $300$~K on PANTHER and MERLIN using the same experimental configuration as for MASS-produced \ce{ZnV2O4} to estimate the phonon scattering contribution. For each sample and incident energy, a vanadium standard was measured to determine the instrumental energy resolution and provide the flux normalisation required to combine spectra collected at different $E_i$ (see below). The instrumental background was measured using an empty aluminium sample holder at each temperature. The quality of the direct background subtraction (without correction for sample attenuation) was assessed from the subtraction of the aluminium Bragg peaks on the elastic line. \\

\noindent While the data collected on PANTHER and LET are presented in the main manuscript, additional PANTHER data and analysis are provided in Supplementary Information Section~\ref{sec:panther}. The corresponding MERLIN data and analysis are presented in Supplementary Information Section~\ref{sec:merlin}. The MERLIN measurements provide an independent check on the systematic accuracy of the PANTHER INS measurements.

\subsubsection*{Inelastic neutron scattering data reduction and phonon estimation}

For all INS experiments, the data were background-subtracted, normalised and reduced using standard scripts in Mantid~\cite{arnoldMantidDataAnalysisVisualization2014}. Following reduction, datasets collected at the same temperature for each measured $E_i$ were combined into a single 3D $S(Q,\Delta E)$ dataset using scaling factors determined from measurements of a vanadium standard at each $E_i$. The combined datasets were then rebinned onto a constant-$Q$ grid with a $Q$-bin width of $0.1$~\text{\AA}$^{-1}$. Equal-temperature data from the combined LET and PANTHER datasets were further combined and rebinned in this way, resulting in combined $S(Q,\Delta E)$ datasets spanning $1\leq Q\leq2$~\text{\AA}$^{-1}$ and $0.4\leq|\Delta E|\leq30$~meV at $1.35\leq Q\leq1.45$~\text{\AA}$^{-1}$. \\

\noindent To extract the inelastic magnetic scattering, $S_\mathrm{mag}(Q,\Delta E)$, estimates of the phonon scattering were calculated and subtracted from the measured INS data. These phonon contributions were estimated from measurements of non-magnetic \ce{ZnGa2O4} collected at 300~K on both PANTHER and MERLIN. It was assumed that phonon contributions at low energy transfers measured on LET were negligible. The phonon density of states, $Z(\Delta E)$, was extracted from the measured phonon spectra using

\begin{equation}
Z(\Delta E)=S_\mathrm{phon}(\Delta E)\frac{\Delta E}{1+n(\Delta E)},
\label{PDOS}
\end{equation} \\

\noindent where $n(\Delta E)$ is the temperature-dependent Bose population factor. The phonon contribution as a function of energy transfer at each constant-$Q$ was estimated by fitting $Z(\Delta E)$ to a back-to-back Gaussian function. For low energy transfers ($|\Delta E|\leq10$~meV on PANTHER and $|\Delta E|\leq14$~meV on MERLIN), where the inelastic phonon signal is obscured by incoherent elastic scattering, the low-energy phonon scattering was approximated by a Debye model, $I\propto\Delta E^2$. The modelled $Z(\Delta E)$ was converted back to $S_\mathrm{phon}(\Delta E)$ for each temperature and $Q$ using Eq.~\ref{PDOS}. The scale factor relating the modelled \ce{ZnGa2O4} phonon scattering to that of \ce{ZnV2O4} was determined by comparison with the \ce{ZnV2O4} $S(Q,\Delta E)$ data collected at $T=300$~K and $5.0<Q<6.0$~\AA$^{-1}$, where the magnetic scattering was assumed to be negligible (see Supplementary Information Section~\ref{sec:INS_phon}).

\subsubsection*{Inelastic neutron scattering data modelling}

\noindent It was found that the magnetic inelastic scattering from MASS-produced \ce{ZnV2O4} could be described by a model consisting of two overdamped harmonic oscillator functions,

\begin{equation}
\chi''_{1,2}(Q,\Delta E)=
\frac{\chi_{1,2}(Q)\Gamma_{1,2}(Q)\Delta E}
{\Gamma_{1,2}^{2}(Q)+\Delta E^{2}}, 
\end{equation} \\

\noindent where $\chi_{1,2}(Q)$ is the $Q$-dependent magnetic susceptibility and $\Gamma_{1,2}(Q)$ is the half-width at half-maximum of each component. Here, $\chi''_{1}$ denotes the low-energy (slowly fluctuating) response and $\chi''_{2}$ the high-energy (fast fluctuating) response. The corresponding $Q$-dependent spin-fluctuation lifetime is given by $\tau_{1,2}(Q)=\hbar/\Gamma_{1,2}(Q)$. The measured scattering intensity is then described by

\begin{equation}
S(Q,\Delta E)=I_{\mathrm{el}}(Q)
+\alpha S_\mathrm{phon}(Q,\Delta E)
+\left[\chi''_{1}(Q,\Delta E)+\chi''_{2}(Q,\Delta E)\right]
\left[1+n(\Delta E)\right].
\label{eq:Scalc}
\end{equation}

\noindent $S_\mathrm{phon}(Q,\Delta E)$ was multiplied by a global scale factor, $\alpha$, during the fitting. The data modelling was performed iteratively, with each dataset treated as follows:

\begin{itemize}

\item The combined LET--PANTHER dataset covers a two-decade energy-transfer range from $0.4\leq|\Delta E|\leq30$~meV at $1.35\leq Q\leq1.45$~\text{\AA}$^{-1}$ and is therefore sensitive to both $\chi''_{1}$ and $\chi''_{2}$. The elastic scattering from the lowest $E_i$ was removed from both the data and the model. Fits to the combined dataset are shown in Fig.~\ref{fig:INS_LP}.

\item The energy resolution of the PANTHER and MERLIN datasets was insufficient to resolve the low-energy component, $\chi''_{1}$, which therefore contributes only to the elastic intensity at these resolutions (see above). Accordingly, this component was fixed to the values obtained from the LET dataset (see below). $I_{\mathrm{el}}(Q)$ was fitted using the measured elastic energy-resolution function from a standard vanadium sample scaled to the data.

\item The LET dataset was sensitive to the low-energy component, $\chi''_{1}$, but did not extend to sufficiently high energy transfers to permit an unconstrained refinement of the high-energy component, $\chi''_{2}$. $\Gamma_{2}$ was therefore fixed to the values obtained from the PANTHER dataset and held constant for each temperature below 50~K. The high-resolution elastic scattering was extracted using the lowest incident energy, $E_i=2.2$~meV and fitted using the scaled elastic energy-resolution function determined from the standard vanadium measurement.

\end{itemize}

\bibliography{ZVO-article}

\newpage

\section*{Data availability}

The raw data generated at the ESRF, ISIS Neutron and Muon Source and Institut Laue-Langevin are available \textit{via} the following links:

\begin{itemize}
    \item \href{https://doi.org/10.15151/ESRF-ES-473795570}{doi:10.15151/ESRF-ES-473795570} (ID22, ESRF),
    \item \href{https://doi.org/10.5286/ISIS.E.RB2090049}{doi:10.5286/ISIS.E.RB2090049} (HRPD, ISIS),
    \item \href{https://doi.org/10.5286/ISIS.E.RB2490091}{doi:10.5286/ISIS.E.RB2490091} (LET, ISIS),
    \item \href{https://doi.org/10.5286/ISIS.E.RB2300055}{doi:10.5286/ISIS.E.RB2300055} (MERLIN, ISIS),
    \item \href{https://doi.org/10.5291/ILL-DATA.DIR-285}{doi:10.5291/ILL-DATA.DIR-285} (PANTHER, ILL),
    \item \href{https://doi.org/10.5291/ILL-DATA.4-03-1783}{doi:10.5291/ILL-DATA.4-03-1783} (PANTHER, ILL).
\end{itemize}

\noindent Data and analysis files supporting the findings of this work, including those associated with the experiments listed above, are available \textit{via} \href{https://doi.org/10.5281/zenodo.19738581}{doi:10.5281/zenodo.19738581}. Further requests for samples, data, analysis files or code should be directed to the corresponding authors.

\section*{Acknowledgements}

This work was supported by the Leverhulme Trust (RPG-2021-366). Part of this work was also supported by the UKRI Engineering and Physical Sciences Research Council (EP/V028774/1). We are grateful to the University of Birmingham for co-funding the PhD studentships of JNG and JLB, in addition to the Institut Laue-Langevin Graduate School (JNG) and the ISIS Facility Development and Utilisation Studentship scheme (JLB). Access to beamtime was funded by the UKRI Science and Technology Facilities Council. We acknowledge Bj\"orn F\aa k (deceased) for local contacting support on PANTHER. We also acknowledge Will Bradbury (LET, ISIS Neutron and Muon Source) and Olivier Meulien (PANTHER, Institut Laue-Langevin) for technical support.

\section*{Author contributions}

LC and JRS conceived the study and supervised all aspects of the work. LC and JAC devised the MASS protocol for \ce{ZnV2O4}, and JAC and JLB scaled-up and optimised the synthesis. JNG, GC and ADF collected the X-ray and neutron diffraction data and JP analysed the data. JLB collected and analysed the SEM data. LC, JRS, JP, SC and JLB collected and analysed the magnetometry data. LC, JRS, JP, SC and DWT collected the INS data, and JRS and JP analysed the data. LC and JRS wrote the manuscript with input from all the authors. LC secured funding for the project. 

\FloatBarrier
\newpage

\begin{appendices}

\renewcommand{\thefigure}{S\arabic{figure}}
\renewcommand{\thetable}{S\arabic{table}}
\setcounter{figure}{0}
\setcounter{table}{0}

\section*{Supplementary Information}\label{ESI}

\subsection{\ce{ZnV2O4} synthesis dependence}\label{sec:PDSI}

\noindent As detailed in the main manuscript, the structural and magnetic properties of \ce{ZnV2O4} are strongly synthesis dependent. Table~\ref{tab:ZVO_Ts} summarises the variation in structural transition temperatures, $T_{\mathrm{S}}$, antiferromagnetic ordering temperatures, $T_{\mathrm{N}}$, and spin-freezing temperatures, $T_{\mathrm{f}}$, reported in the literature. \\

\begin{table}[hbt!]
    \caption{Reported spin-freezing temperatures, $T_{\mathrm{f}}$, antiferromagnetic ordering temperatures, $T_{\mathrm{N}}$, and cubic-to-tetragonal structural transition temperatures, $T_{\mathrm{S}}$, for \ce{ZnV2O4} prepared using different synthesis routes.}
    \begin{tabular}{lcccc}
        \toprule
         & Synthesis & $T_{\mathrm{f}}$~(K) & $T_{\mathrm{N}}$~(K) & $T_{\mathrm{S}}$~(K) \\
        \midrule
        Nizioł \textit{et al.}\cite{niziolInvestigationMagneticProperties1973} & Furnace & -- & 45 & -- \\
        Ueda \textit{et al.}\cite{uedaMagneticStructuralTransitions1997} & Furnace & -- & 40 & 50 \\
        Nishiguchi \textit{et al.}\cite{nishiguchiPseudotetramerGeometricallyFrustrated2002} & Furnace & -- & 44 & 52 \\
        Reehuis \textit{et al.}\cite{reehuisCrystallographicMagneticStructure2003} & Furnace & -- & 40 & 51 \\
        Lee \textit{et al.}\cite{leeOrbitalSpinChains2004} & Furnace & -- & 40 & 50 \\
        Vasiliev \textit{et al.}\cite{vasilievSpecificHeatMagnetic2006} & Furnace & -- & 31 & 45 \\
        Blanco-Canosa \textit{et al.}\cite{blanco-canosaEnhancedPressureDependence2007} & Furnace & -- & -- & 38 \\
        Wang \textit{et al.}\cite{wangPhaseTransitionCritical2015} & Furnace & -- & 40 & 50 \\
        Singha \textit{et al.}\cite{singhaStructuralMagneticStudies2019} & Furnace & -- & 41 & 57 \\
        Browne \textit{et al.}\cite{browneOrbitalMoleculesVanadium2020} & Furnace & -- & 39 & 50 \\
        Ebbinghaus \textit{et al.}\cite{ebbinghausCrystalStructureMagnetic2004} & CVT & 11 & -- & -- \\
        Shrivastava \textit{et al.}\cite{shrivastavaModulatingOpticalMagnetic2020} & Sol-gel & 13 & -- & -- \\
        \bottomrule
    \end{tabular}
    \label{tab:ZVO_Ts}
\end{table}

\FloatBarrier
\newpage


\subsection{Optimisation and reproducibility of the MASS method}\label{sec:MASS_opt}

\begin{figure}[hbt!]
    \centering
    \includegraphics[width=0.95\textwidth]{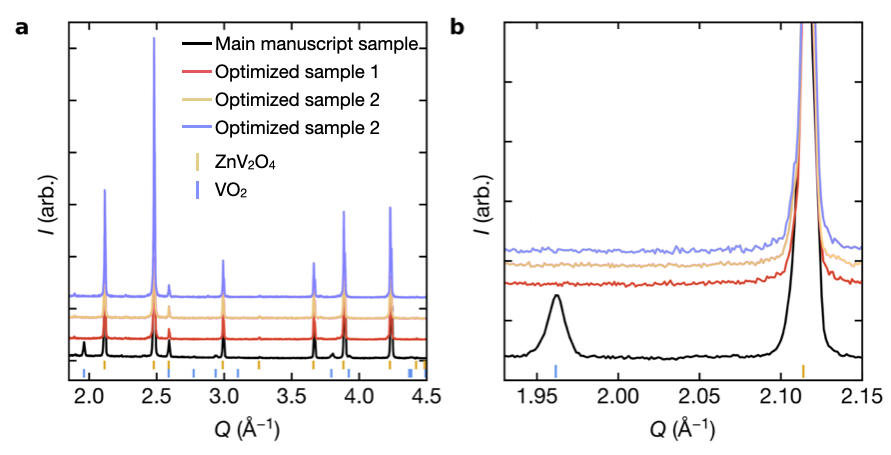}
    \caption{Laboratory powder X-ray diffraction data for three independently prepared optimised MASS-produced \ce{ZnV2O4} samples over \textbf{a} $1.8<Q<4.5$~\AA$^{-1}$ and \textbf{b} $1.93<Q<2.15$~\AA$^{-1}$ (shown on a $5\times$ expanded intensity scale), demonstrating that optimisation of the MASS synthesis reproducibly reduces the \ce{VO2} impurity phase below the detection limit of laboratory powder X-ray diffraction.}
    \label{fig:synth_dep_2}
\end{figure}

\noindent To minimise the \ce{VO2} impurity phase present in the initial MASS-produced samples of \ce{ZnV2O4}, the synthesis procedure was further optimised as described in the Methods. Figure~\ref{fig:synth_dep_2} shows laboratory powder X-ray diffraction data for three independently prepared optimised MASS samples. The \ce{VO2} impurity phase is reduced below the detection limit of laboratory powder X-ray diffraction in all three samples, demonstrating the reproducibility of the optimised synthesis procedure. \\

\begin{figure}[hbt!]
    \centering
    \includegraphics[width=0.75\textwidth]{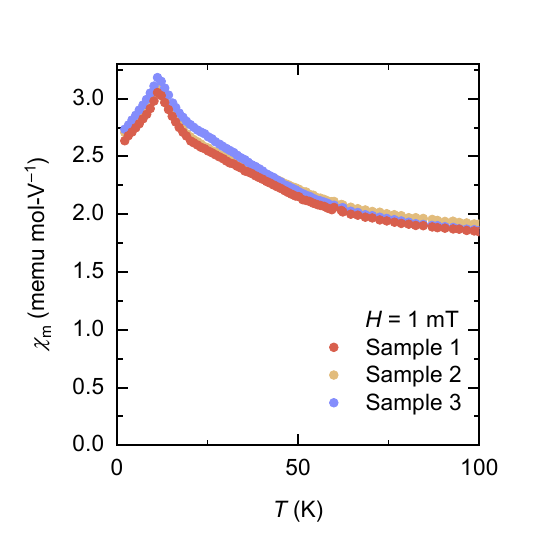}
    \caption{Temperature dependence of the molar magnetic susceptibility, $\chi_{\mathrm{m}}$, of three independently prepared optimised MASS-produced \ce{ZnV2O4} samples. All samples exhibit the same characteristic magnetic behaviour, including the sharp cusp at $T\approx11$~K, the broader feature at $T\approx40$~K, and the absence of the structural ($T_{\mathrm{S}}$) and antiferromagnetic ($T_{\mathrm{N}}$) transitions characteristic of conventional furnace-prepared \ce{ZnV2O4}. These data demonstrate the reproducibility of the optimised MASS synthesis and show that the characteristic magnetic phenomenology of MASS-produced \ce{ZnV2O4} is retained following optimisation of the synthesis procedure to reduce the \ce{VO2} impurity phase below the detection limit of laboratory powder X-ray diffraction.}
    \label{fig:synth_dep_3}
\end{figure}

\noindent To establish whether removal of the \ce{VO2} impurity alters the intrinsic magnetic behaviour of MASS-produced \ce{ZnV2O4}, Fig.~\ref{fig:synth_dep_3} compares the magnetic susceptibility of three independently prepared optimised samples. All samples exhibit the same characteristic magnetic response, including the pronounced cusp in $\chi_{\mathrm{m}}$ at $T\approx11$~K and the broader feature at $T\approx40$~K, together with the absence of the structural ($T_{\mathrm{S}}$) and antiferromagnetic ($T_{\mathrm{N}}$) transitions characteristic of conventional furnace-prepared samples. The persistence of these features following removal of the \ce{VO2} impurity demonstrates that the \ce{VO2} impurity present in the sample discussed in the main manuscript does not measurably influence the characteristic magnetic phenomenology. Moreover, their observation across independently prepared samples demonstrates that the MASS method reproducibly stabilises the same distinct magnetic ground-state behaviour. \\

\noindent We note, however, that the absolute magnitude of $\chi_{\mathrm{m}}$ is larger for the optimised samples than for the DC susceptibility data presented in the main text. The absolute susceptibility values should be compared with caution for two reasons. First, the data are normalised to the nominal amount of \ce{ZnV2O4}. For samples containing a secondary phase, the actual mass of \ce{ZnV2O4} is lower than the total measured sample mass; normalisation using the latter therefore underestimates the molar susceptibility of the \ce{ZnV2O4} phase. Second, the magnitude of the susceptibility of \ce{ZnV2O4} is strongly dependent on applied magnetic field, particularly at the low fields used for these measurements. Small differences in field nulling and residual magnetic field between measurements may therefore also contribute to variations in the absolute susceptibility. These uncertainties affect quantitative comparison of the absolute magnitude of $\chi_{\mathrm{m}}$, but not the reproducible temperature-dependent features used here to identify the magnetic ground-state phenomenology.

\FloatBarrier
\pagebreak

\subsection{Rietveld analysis of high-resolution diffraction data}\label{sec:RASI}

\noindent We tested for site vacancies and cation site mixing in MASS-produced \ce{ZnV2O4} by simultaneous Rietveld refinement against the high-resolution synchrotron powder X-ray and neutron powder diffraction data. Using the minimal structural model described in the main manuscript as the starting point, we explored refinement of the occupancies of the Zn$^{2+}$, V$^{3+}$ and O$^{2-}$ sites, individually and in combination (Tab.~\ref{tab:riet_1}). We also tested models incorporating Zn$^{2+}$/V$^{3+}$ site mixing with both fixed and refined O$^{2-}$ occupancies (Tab.~\ref{tab:riet_2}). \\

\begin{figure}[hbt!]
    \centering
    \includegraphics[width=0.95\textwidth]{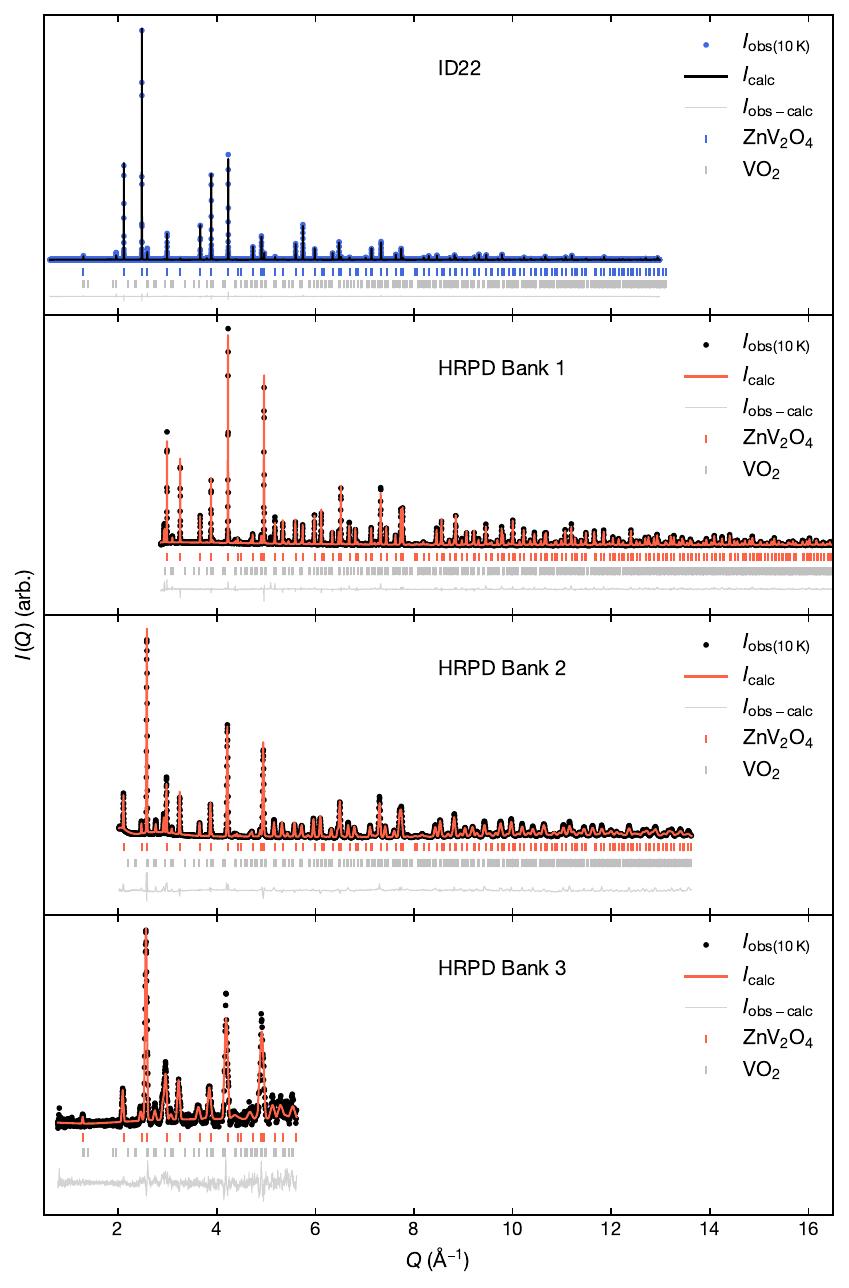}
    \caption{Rietveld refinement of the \ce{ZnV2O4} cubic spinel phase at $T=10$~K (minimal model) and \ce{VO2} impurity phase against powder X-ray diffraction data collected on ID22 at the ESRF \cite{fitchID22HighresolutionPowderdiffraction2023} and powder neutron diffraction data collected on Banks 1, 2 and 3 of HRPD at ISIS Neutron and Muon Source \cite{ibbersonDesignPerformanceNew2009}.}
    \thisfloatpagestyle{empty}
    \label{fig:Riet}
\end{figure}

\noindent Although refinement of the site occupancies leads to modest improvements in the fit metrics, the resulting models require unrealistically high concentrations of oxygen vacancies that would necessitate reduction of V$^{3+}$ to V$^{2+}$ for charge balance, which is chemically implausible. Similarly, refinement of Zn$^{2+}$/V$^{3+}$ site mixing drives the fit into a local minimum and results in poorer overall agreement with the diffraction data (Tab.~\ref{tab:riet_2}). We therefore find no chemically plausible model involving vacancies or cation site disorder that provides an improvement over the minimal structural model. Corresponding Rietveld refinement plots for the minimal model are shown in Fig.~\ref{fig:Riet}.

\FloatBarrier
\newpage

\begin{sidewaystable}[hbt!]
        \caption{Rietveld refinement parameters from the minimal model and the further refinements of \ce{Zn}, \ce{V} and \ce{O} site occupancies, both individually and in combination. The atomic displacement parameters were fixed to those obtained from the minimal model.}
\begin{tabular}{lllllllll}
    \toprule
    Parameter & Minimal model & Zn Occ. & V Occ. & O Occ. & Zn \& O Occ. & V \& O Occ. & Zn \& V Occ. & Zn, V \& O Occ.\\
    \midrule
    $a$ ($\mathrm{\text{\AA}}$) & 8.39638(1) & 8.39638(1) & 8.39638(1) & 8.39638(1) & 8.39638(1) & 8.39638(1) & 8.39638(1) & 8.39638(1)  \\
    $\epsilon$ & 0.000197(3) & 0.000197(3) & 0.000197(3) & 0.000197(3) & 0.000196(3) & 0.000197(3) & 0.000197(3) & 0.000196(3) \\
    Crystallite size (nm) & 340(2) & 338(2) & 340(2) & 341(2) & 339(2) & 340(2) & 339(2) & 339(2) \\
    Zn $B_{iso}$ & 0.212(5) & 0.212 & 0.212 & 0.212 & 0.212 & 0.212 & 0.212 & 0.212 \\
    V $B_{iso}$ & 0.135(4) & 0.135 & 0.135 & 0.135 & 0.135 & 0.135 & 0.135 & 0.135 \\
    O $B_{iso}$ & 0.462(13) & 0.462 & 0.462 & 0.462 & 0.462 & 0.462 & 0.462 & 0.462 \\
    Zn Occ. & 1 & 0.9791(9) & 1 & 1 & 0.9724(1) & 1 & 0.985(2) & 0.9727(6) \\
    V Occ. & 1 & 1 & 1.000(1) & 1 & 1 & 1.000(1) & 1.000(2) & 1.0000(6) \\
    O Occ. & 1 & 1 & 1 & 0.977(2) & 0.957(2) & 0.980(2) & 1 & 0.957(2) \\
    O Coordinate & 0.23927(6) & 0.23945(6) & 0.23936(6) & 0.23931(6) & 0.23965(6) & 0.23937(6) & 0.23949(6) & 0.23965(6) \\
    ID22 $R_{wp}$ & 13.143800 & 13.042680 & 13.157670 & 13.118620 & 12.961550 & 13.125420 & 13.061810 & 12.961630 \\
    Bank 1 $R_{wp}$ & 5.473560 & 5.266670 & 5.351850 & 5.572670 & 5.353970 & 5.505690 & 5.246830 & 5.350970 \\
    Bank 2 $R_{wp}$ & 5.613810 & 5.584440 & 5.568720 & 5.605810 & 5.554770 & 5.585700 & 5.564650 & 5.552220 \\
    Bank 3 $R_{wp}$ & 9.413900 & 9.399690 & 9.373800 & 9.402170 & 9.370260 & 9.391890 & 9.376430 & 9.359110 \\
    GOF & 2.557540 & 2.537770 & 2.560150 & 2.552560 & 2.522030 & 2.553920 & 2.541530 & 2.522090 \\
    $R_{wp}$ & 13.139470 & 13.038350 & 13.153300 & 13.114340 & 12.957280 & 13.121110 & 13.057460 & 12.957360 \\
    \bottomrule
    \end{tabular}
    \label{tab:riet_1}
\end{sidewaystable}

\newpage

\begin{table}[hbt!]
        \caption{Rietveld refinement parameters from the minimal model and the further refinements of \ce{Zn}/\ce{V} site mixing, both with and without freely refining the \ce{O} site occupancies. The atomic displacement parameters were fixed to those obtained from the minimal model.}
\begin{tabular}{llll}
    \toprule
    Parameter & Minimal model & \ce{Zn}/\ce{V} site mixing & \ce{Zn}/\ce{V} site mixing + O Occ. \\
    \midrule
    $a$ ($\mathrm{\text{\text{\AA}}}$) & 8.39638(1) & 8.39639(1) & 8.39639(1) \\
    $\epsilon$ & 0.000197(3) & 0.000184(3) & 0.000184(3) \\
    Crystallite size (nm) & 340(2) & 338(2) & 338(2) \\
    Zn $B_{iso}$ & 0.212(5) & 0.212 & 0.212 \\
    V $B_{iso}$ & 0.135(4) & 0.135 & 0.135 \\
    O $B_{iso}$ & 0.462(13) & 0.462 & 0.462 \\
    Zn Occ. & 1 & 0.931(2) & 0.929(2) \\
    V Occ. & 1 & 0.965(1) & 0.965(1) \\
    O Occ. & 1 & 1 & 0.989(2) \\
    O Coordinate & 0.23927(6) & 0.23970(6) & 0.23974(6) \\
    ID22 $R_{wp}$ & 13.14380 & 13.60718 & 13.60198 \\
    Bank 1 $R_{wp}$ & 5.47356 & 4.24343 & 4.18765 \\
    Bank 2 $R_{wp}$ & 5.61381 & 5.09705 & 5.05663 \\
    Bank 3 $R_{wp}$ & 9.41390 & 9.18026 & 9.33272 \\
    GOF & 2.557540 & 2.647540 & 2.64657 \\
    $R_{wp}$ & 13.139470 & 13.602310 & 13.59710 \\
    \bottomrule
    \end{tabular} 
    \label{tab:riet_2}
\end{table}

\FloatBarrier
\pagebreak


\subsection{Scanning electron microscopy and particle size analysis}

\begin{figure}[hbt!]
    \centering
    \includegraphics[width=0.95\textwidth]{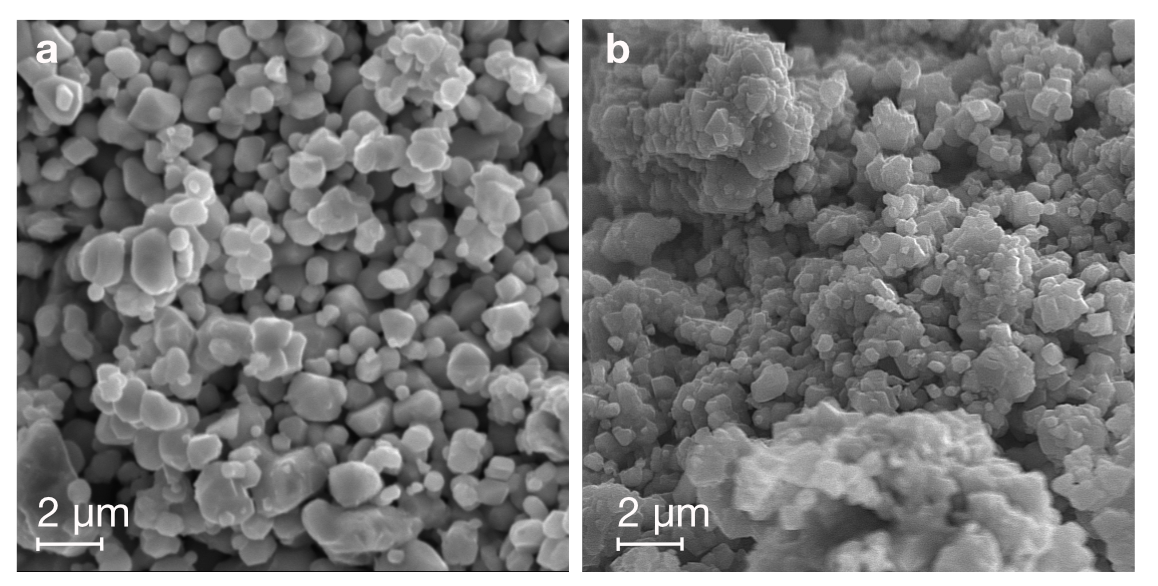}
    \caption{SEM micrographs of \textbf{a} MASS-produced and \textbf{b} furnace-prepared \ce{ZnV2O4}.}
    \label{fig:SEM}
\end{figure}

\begin{figure}[hbt!]
    \centering
    \includegraphics{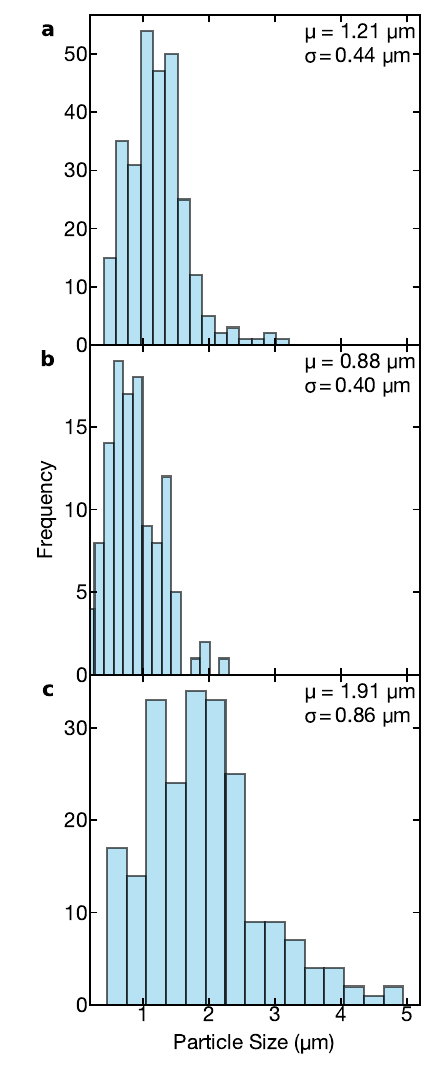}
    \caption{Particle diameter distributions determined from SEM images of \textbf{a} the MASS-produced sample investigated in the main manuscript, \textbf{b} a furnace-prepared sample, and \textbf{c} an optimised MASS-produced sample. The mean particle diameter, $\mu$, and standard deviation, $\sigma$, obtained from the measured distributions are indicated in each panel.}
    \label{fig:PSA}
\end{figure}

\noindent To compare the particle morphology of MASS-produced \ce{ZnV2O4} with that of a typical furnace-prepared sample, SEM images were collected for each sample (see Methods), as shown in Fig.~\ref{fig:SEM}. Particle size analysis indicates a larger average particle diameter for the MASS-produced sample, $1.21(44)$~$\mu$m, than for the furnace-prepared sample, $0.88(40)$~$\mu$m (Fig.~\ref{fig:PSA}). The optimised MASS-produced sample also exhibits a comparable particle size distribution, indicating that optimisation of the synthesis procedure does not substantially alter the particle morphology.

\FloatBarrier
\pagebreak

\subsection{DC susceptibility analysis}

\noindent The Weiss constant, $\theta$, and Curie constant, $C$, were determined from the temperature-dependent magnetic susceptibility of MASS-produced \ce{ZnV2O4} using a Husimi ansatz \cite{pohleCurielawCrossoverSpin2023},

\begin{equation}
\chi_\mathrm{m}T(T)=\frac{1+b_1\exp(c_1/T)}{a+b_2\exp(c_2/T)}.
\label{eq:HA}
\end{equation} \\

\begin{figure}[hbt!]
\centering
\includegraphics{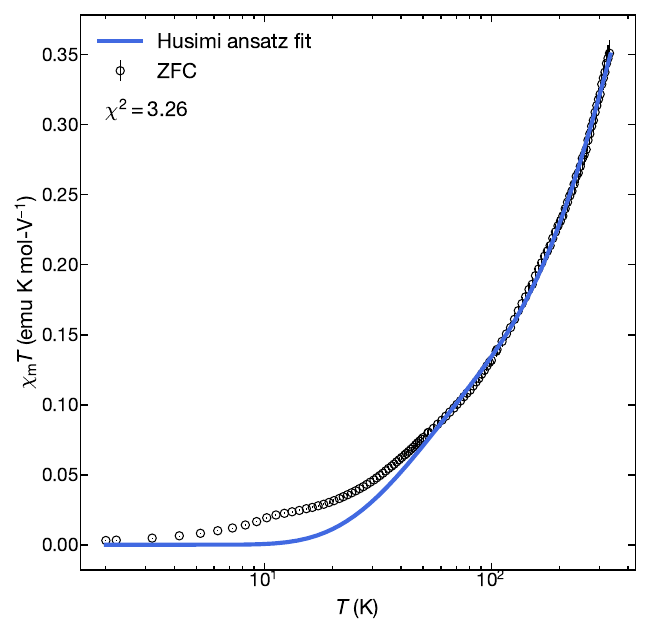}
\caption{Reduced DC susceptibility, $\chi_{\mathrm{m}}T(T)$, of MASS-produced \ce{ZnV2O4} measured in a $1$~mT magnetic field under zero-field-cooled (ZFC) conditions together with the Husimi ansatz fit over the temperature range $50<T<335$~K~\cite{pohleCurielawCrossoverSpin2023}.}
\label{fig:DC_chiT}
\end{figure}

\noindent The Husimi ansatz is applicable within the correlated paramagnetic regime ($T<|\theta|$), where the magnetic susceptibility deviates from Curie--Weiss behaviour, and has recently been shown to provide an accurate description of the $S=1$ pyrochlore antiferromagnet \ce{NaCaNi2F7}~\cite{pohleCurielawCrossoverSpin2023}. The reduced susceptibility, $\chi_\mathrm{m}T(T)$, was fitted to Eq.~\ref{eq:HA} over the temperature range $50<T<335$~K using least-squares minimisation (Fig.~\ref{fig:DC_chiT}). During unconstrained refinement, the parameter $a$ was found to be zero within experimental uncertainty and was therefore fixed at zero for the final refinement. The Curie and Weiss constants were then obtained from the refined parameters using

\begin{equation}
C=\frac{1+b_1}{a+b_2},
\qquad
\theta=\frac{b_1c_1}{1+b_1}-\frac{b_2c_2}{a+b_2}.
\end{equation} \\

\FloatBarrier
\newpage


\subsubsection{AC susceptibility analysis of the frequency dependence of $T_1$ and $T_2$}
\label{sec:AC}

\begin{figure}[hbt!]
    \centering
    \includegraphics[width=0.7\textwidth]{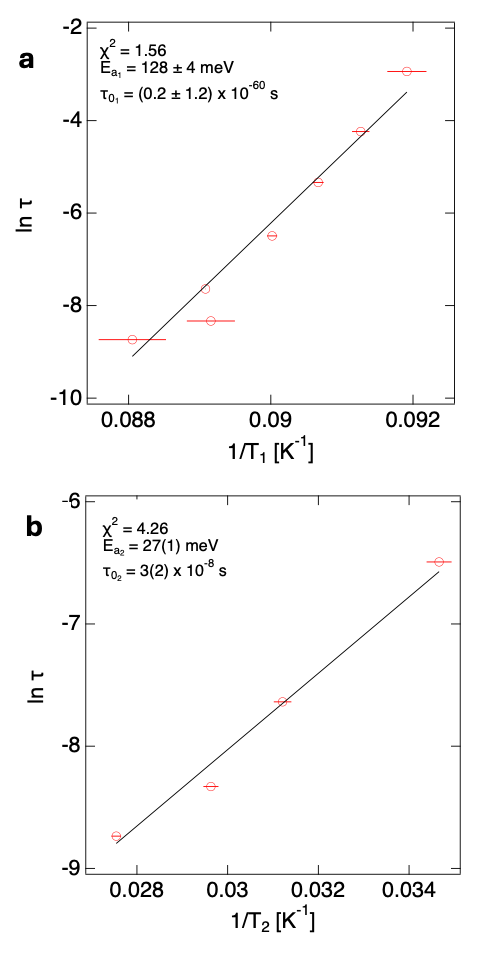}
    \caption{$\ln \tau$ as a function of $1/T_{\mathrm{f}}$ (red circles), where $T_{\mathrm{f}} = T_1$ {\bf a} and $T_{\mathrm{f}} = T_2$ {\bf b}. Both regimes are modelled using the Arrhenius form, Eq.~\ref{eq:Arrhenius} (black solid line), from which the characteristic spin-relaxation times, $\tau_0$, and activation energies, $E_a$, are determined.}
    \label{fig:AC_chi_1}
    \thisfloatpagestyle{empty}
\end{figure}

The spin-freezing temperature of \ce{ZnV2O4}, $T_1$, was determined by fitting the real part of the AC susceptibility data, $\chi'$, to a back-to-back Gaussian function for all measured frequencies. The extracted $T_1$ values are shown in Fig.~\ref{fig:ACDC_chi}c as a function of frequency. In Fig.~\ref{fig:AC_chi_1}a, the logarithm of the spin-relaxation time, $\ln\tau = \ln\left(\frac{1}{2\pi f_{\mathrm{AC}}}\right)$, is plotted as a function of the inverse freezing temperature and fitted to an Arrhenius form,

\begin{equation}
\tau = \tau_0~\exp\bigg(\frac{E_a}{k_BT_{\mathrm{f}}}\bigg) \Rightarrow \ln\tau = \ln\tau_0 + \bigg(\frac{E_a}{k_B}\bigg)\bigg(\frac{1}{T_{\mathrm{f}}}\bigg)
\label{eq:Arrhenius}
\end{equation}\\

\noindent where $T_{\mathrm{f}}$ is the freezing temperature, $\tau_0$ is the characteristic relaxation time of the frozen spins and $E_a$ is the Arrhenius activation energy barrier to flipping a frozen spin. We find that $\ln\tau$ follows activated behaviour with $E_{a_1} = (128 \pm 4)$~meV and a vanishingly small (zero within error) characteristic spin-relaxation time, $\tau_0$. \\

\noindent The broad frequency-dependent peak seen in $\chi''_{ac}$ for $f \geq 105$ Hz is fitted to Gaussian curves (a back-to-back Gaussian for the more asymmetric 105 Hz data) to extract the high-temperature peak $T_2$ values (Table~\ref{tab:Tvstau}). 

\begin{table}[h!]
\centering
\caption{$T_2$ and $\tau$ values extracted from $\chi_{ac}''(T)$} 
\label{tab:Tvstau}
\begin{tabular}{l c c c}
\hline
\hline
$f$ (Hz) & $\tau = (2\pi f)^{-1}$ (s) & $\ln \tau$ & $T_2 \pm \delta T_2$ \\
\hline
990 & 1.608 $\times 10^{-4}$ & -8.73558 & 36.30822 $\pm 0.14369$\\
660 & 2.411 $\times 10^{-4}$ & -8.33012 & 33.74581 $\pm 0.18794$\\
330 & 4.823 $\times 10^{-4}$ & -7.63697 & 32.04139 $\pm 0.20444$\\
105 & 1.516 $\times 10^{-3}$ & -6.49184 & 28.84931 $\pm 0.23233$\\
\hline
\hline
\end{tabular}
\end{table}

\noindent In Fig.~\ref{fig:AC_chi_1}b, $\ln\tau$ vs $1/T_2$ is again shown to be well described by an Arrhenius form (Eq.~\ref{eq:Arrhenius}). Here, we find a much lower activation energy, $E_{a_2} = (27 \pm 1)$~meV, and a longer characteristic spin-relaxation time of $\tau_{0_2} = (30 \pm 20)$~ns.

\FloatBarrier
\newpage

\subsection{Temperature dependence of the magnetic correlation length}
\label{sec:corrs}

\noindent The real-space extent of the magnetically correlated high-energy spin-fluctuations in \ce{ZnV2O4}  can be extracted as a function of temperature from the $Q$- and $T$-dependence of the static susceptibility measured on PANTHER, $\chi_2(Q,T)$, as shown in Fig.~\ref{fig:INS_metrics}c.  Assuming exponential decay of spin correlations in real space, this was fitted to a Lorentzian form,
\[\chi_2(Q,T) \propto \frac{1}{1/\xi^2(T) + (Q-Q_\mathrm{AF}(T))^2} + C\]
where $\xi(T)$ is the spin correlation range and $Q_\mathrm{AF}(T)$ is the peak centre, which is related to the antiferromagnetic interaction distance, $r_\mathrm{AF}(T)$ (see below).  $C$ is a temperature and $Q$-independent background which takes account of any unsubtracted phonons or incoherent scattering beneath the extracted magnetic intensity. $C$ was estimated from the $T=300$~K data set where, by inspection, $\chi_2(Q)$ is independent of $Q$. In the presence of purely first near-neighbour antiferromagnetic correlations, the magnetic short-range order scattering takes the form of a so-called {\it dimer structure factor},
\[I(Q) \propto 1-\frac{\sin(Qr_\mathrm{AF})}{Qr_\mathrm{AF}}.\]
The value of $r_\mathrm{AF}$ corresponding to $Q_\mathrm{AF}$ is found by differentiating $I(Q)$ with respect to $Q$ and then setting to zero,
\begin{eqnarray*}
I'(Q) &=& \frac{Q_\mathrm{AF}r_\mathrm{AF}^2\cos(Q_\mathrm{AF}r_\mathrm{AF}) - r_\mathrm{AF}\sin(Q_\mathrm{AF}r_\mathrm{AF})}{Q_\mathrm{AF}^2r_\mathrm{AF}^2}=0 \\
&\Rightarrow& Q_\mathrm{AF}r_\mathrm{AF} = \tan(Q_\mathrm{AF}r_\mathrm{AF}).
\end{eqnarray*}
This can be solved numerically or graphically, where the roots are the turning points of the dimer structure factor.  The first positive non-zero solution corresponding to the position of the first peak at $Q_\mathrm{AF}$ is given by,
\[Q_\mathrm{AF}r_\mathrm{AF} \simeq 4.5\]
from which $r_\mathrm{AF}(T)$ may be calculated at each temperature.\\

\begin{figure}[hbt!]
    \centering
    \includegraphics{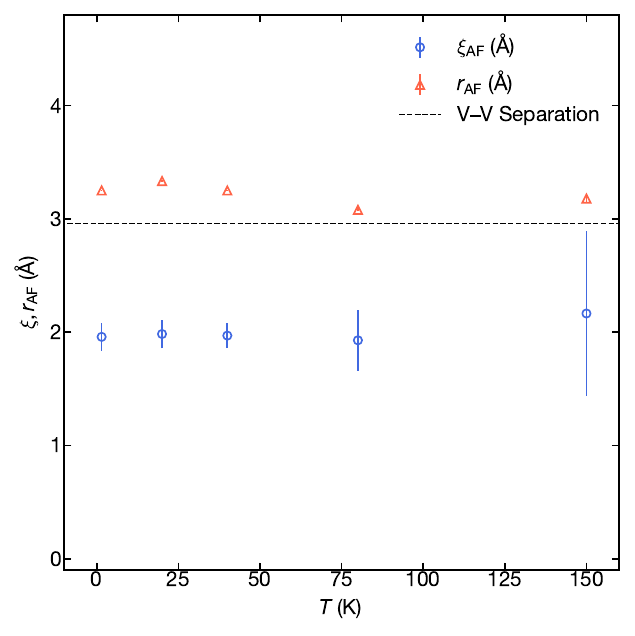}
    \caption{The temperature dependence of the near-neighbour antiferromagnetic interaction distance, $r_\mathrm{AF}(T)$, and the antiferromagnetic correlation range, $\xi_\mathrm{AF}(T)$. The V-V near-neighbour distance on the pyrochlore sublattice is indicated by a dashed line.}
    \label{fig:chi2_lor}
\end{figure}

\noindent $\xi(T)$ and $r_\mathrm{AF}(T)$ are plotted in Fig.~\ref{fig:chi2_lor} as a function of temperature between $1.5 \le T \le 150$~K and are, remarkably, independent of temperature over this range.\\

\noindent This indicates that the extent of the spin correlations is limited by magnetic geometrical frustration and quantum---\textit{not} thermal---fluctuations, never developing short-range order beyond first V-V near-neighbours. In addition, the data indicate that in MASS-prepared \ce{ZnV2O4}, the magnetic correlations are fully three-dimensional, with no indication of an asymmetrical lineshape as would be expected for one-dimensional chain correlations, as was observed in the neutron scattering experiments of Lee and co-workers \cite{leeOrbitalSpinChains2004}.\\

\FloatBarrier
\newpage

\subsection{Phonon inelastic neutron scattering estimation on PANTHER}
\label{sec:INS_phon}
\begin{figure}[hbt!]
    \centering
    \includegraphics{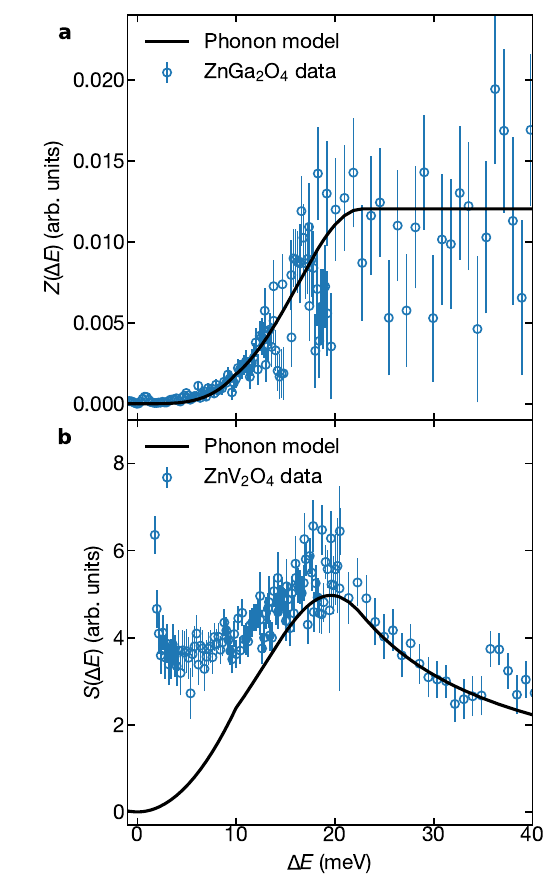}

    \caption{\textbf{a} The observed INS data obtained from \ce{ZnGa2O4} integrated over $5.55 \le Q \le 5.65$~\AA$^{-1}$ and $T = 300\;\mathrm{K}$, converted to $Z(\Delta E)$ and fitted by a back-to-back Gaussian function. \textbf{b} The observed INS data obtained from MASS-prepared \ce{ZnV2O4} integrated over $5.55 \le Q \le 5.65$~\AA$^{-1}$ and measured at $T = 150\;\mathrm{K}$ and the phonon model, which was extracted from $Z(\Delta E)$ using Eq.~\ref{PDOS} and scaled to the data, as described in the Methods section of the Main Manuscript.}
    \label{fig:INS_phon}
\end{figure}

The estimated phonon contribution as a function of energy for the constant-$Q$ cut at $5.55 \le Q \le 5.65$~\AA$^{-1}$ is shown in Fig. \ref{fig:INS_phon}. In Fig. \ref{fig:INS_phon}a, we show the observed INS data obtained from \ce{ZnGa2O4}, converted to the phonon density of states, $Z(\Delta E)$, and modelled using a back-to-back Gaussian function. The lower energy side of the back-to-back Gaussian captures $Z(\Delta E)$ at $\Delta E \lesssim 20~$meV, while the higher energy side of the back-to-back Gaussian captures the near constant $Z(\Delta E)$ at $\Delta E \gtrsim 20~$meV. Fig. \ref{fig:INS_phon}b shows the phonon estimation, $S_\mathrm{phon}(\Delta E)$, obtained by converting the fitted phonon model to $S(\Delta E)$ (Eq.~\ref{PDOS}), replacing it with a Debye model ($S \propto \Delta E^2$) at $\Delta E \le 10~$meV, where the data are contaminated by the elastic scattering resolution function, and scaling it to the observed INS data obtained from \ce{ZnV2O4}, $S_\mathrm{exp}(\Delta E)$. The function displayed in Fig. \ref{fig:INS_phon}b is then included in the model shown in Fig.~\ref{fig:INS_P} with a refined scale factor.

\FloatBarrier
\newpage

\subsection{Inelastic neutron scattering modelling on PANTHER}
\label{sec:panther}
\begin{figure}[hbt!]
    \centering
    \includegraphics[width=1\textwidth]{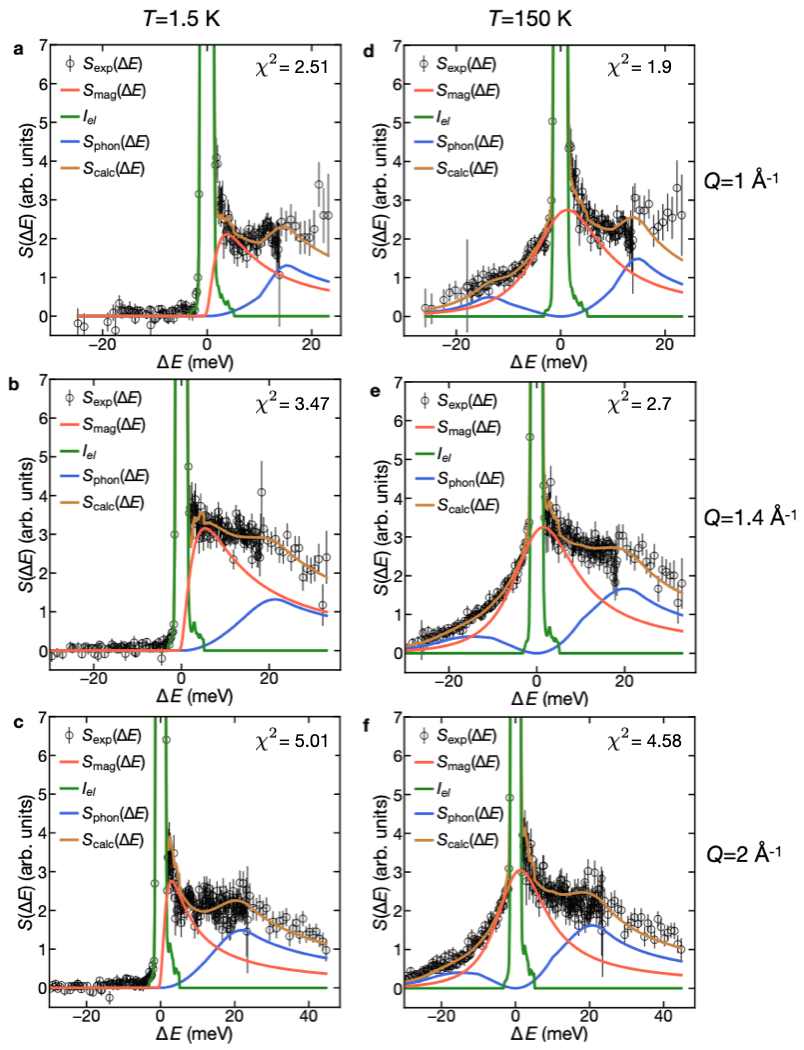}
    \caption{The observed INS data, $S_\mathrm{exp}(\Delta E)$, fitted inelastic magnetic scattering, $S_\mathrm{mag}(\Delta E)$, elastic line profile, $I_\mathrm{el}$,$^\dag$ phonon model, $S_\mathrm{phon}(\Delta E)$, and total calculated function, $S_\mathrm{calc}(\Delta E)$, for the PANTHER $S_\mathrm{exp}(\Delta E)$ at $T = \left\{1.8, 150\right\}$~K.  Data were integrated over $Q$-ranges of:     
    \textbf{a} $0.95 \le Q \le 1.05$~\AA$^{-1}$, 
    \textbf{b} $1.35 \le Q \le 1.45$~\AA$^{-1}$, 
    \textbf{c} $1.95 \le Q \le 2.05$~\AA$^{-1}$, 
    \textbf{d} $0.95 \le Q \le 1.05$~\AA$^{-1}$, 
    \textbf{e} $1.35 \le Q \le 1.45$~\AA$^{-1}$, 
    \textbf{f} $1.95 \le Q \le 2.05$~\AA$^{-1}$. 
    Reduced-$\chi^2$ goodness-of-fit parameters are shown in the top-tight of each panel. $^\dag$$I_\mathrm{el}$ was fitted using the measured elastic energy-resolution function from a standard vanadium sample scaled to the data.}
    \label{fig:INS_P}
    \thisfloatpagestyle{empty}
\end{figure}

Examples of modelling of constant-$Q$ cuts at $0.95 \le Q \le 1.05$,~$1.35 \le Q \le 1.45$~and~$1.95 \le Q \le 2.05$~\AA$^{-1}$ of the PANTHER data at $T=1.5$ and $150$~K are presented in Fig.~\ref{fig:INS_P}. The data are modelled according to Equation~\ref{eq:Scalc}.

\FloatBarrier
\newpage

\subsection{Inelastic neutron scattering modelling on MERLIN}
\label{sec:merlin}

\begin{figure}[hbt!]
    \centering
    \includegraphics{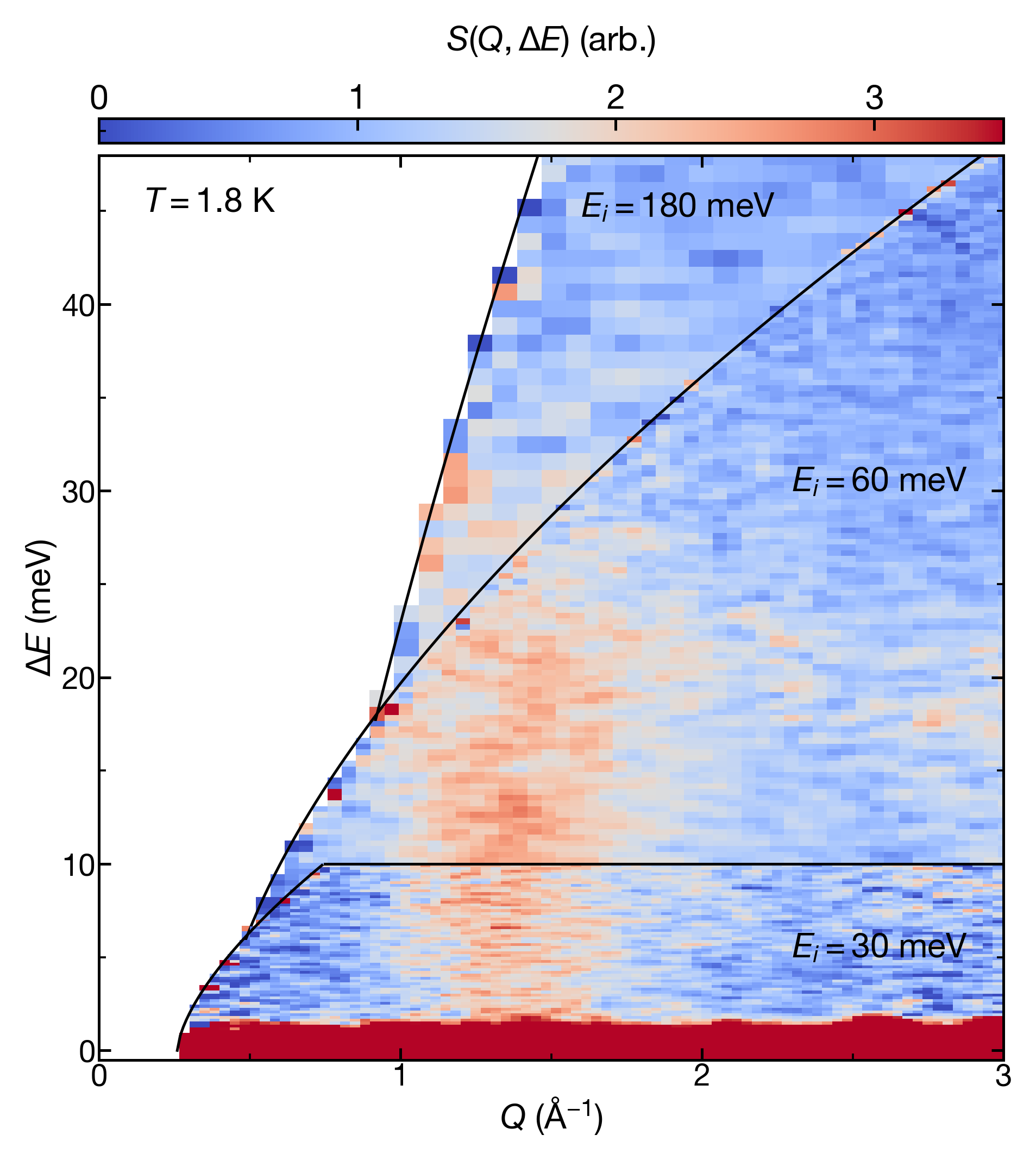}
        \caption{The dynamical structure factor, $S(Q,\Delta E)$, of MASS-produced \ce{ZnV2O4} obtained by INS measurement on the MERLIN spectrometer at $T=1.8$~K and $E_i=30$, $90$ and $180$~meV.}
    \label{fig:merlin_slices}
\end{figure}

\begin{figure}[hbt!]
    \centering
    \includegraphics[width=1\textwidth]{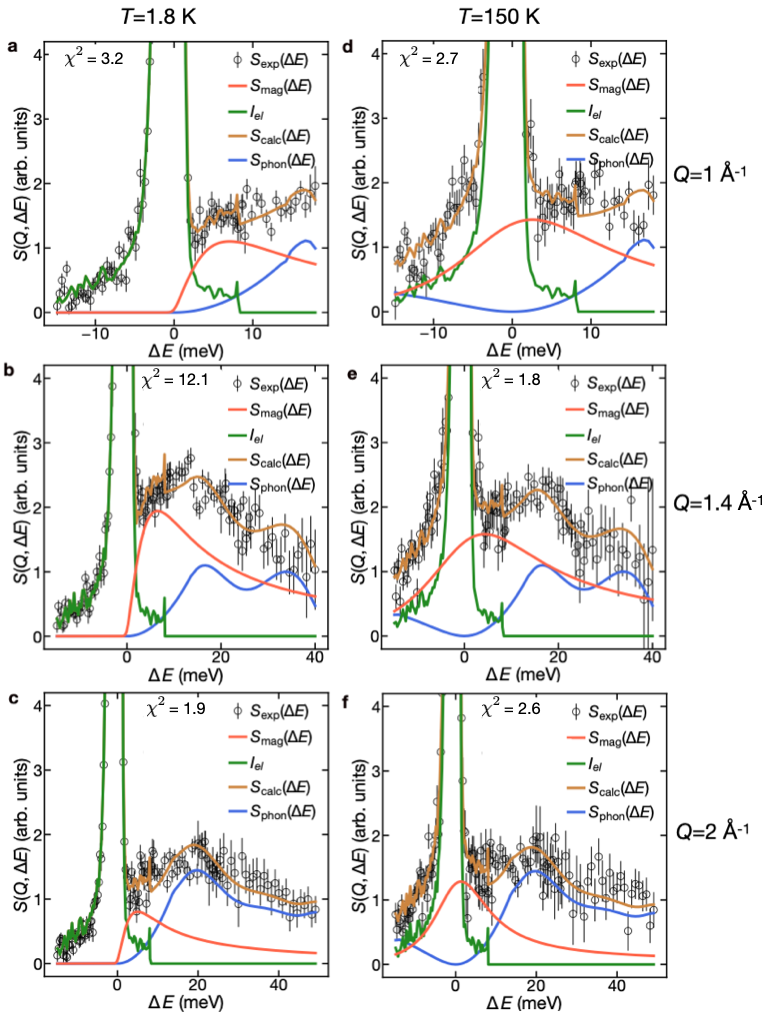}
    \caption{The observed INS data, $S(\Delta E)$, fitted inelastic magnetic scattering, $S_\mathrm{mag}(\Delta E)$, elastic line profile, $I_\mathrm{el}$,$^\dag$ phonon model, $S_\mathrm{phon}(\Delta E)$, and total calculated function, $S_\mathrm{calc}(\Delta E)$, for the MERLIN $S(\Delta E)$ at $T = \{1.8, 150\}$~K. Data shown were integrated over $Q$-ranges of: 
    \textbf{a} $0.95 \le Q \le 1.05$~\AA$^{-1}$, 
    \textbf{b} $1.35 \le Q \le 1.45$~\AA$^{-1}$, 
    \textbf{c} $1.95 \le Q \le 2.05$~\AA$^{-1}$, 
    \textbf{d} $0.95 \le Q \le 1.05$~\AA$^{-1}$, 
    \textbf{e} $1.35 \le Q \le 1.45$~\AA$^{-1}$, 
    \textbf{f} $1.95 \le Q \le 2.05$~\AA$^{-1}$. 
    Reduced-$\chi^2$ goodness-of-fit parameters are shown at the top of each panel. $^\dag$$I_\mathrm{el}$ was fitted using the measured elastic energy-resolution function from a standard vanadium sample scaled to the data.}
    \label{fig:INS_M}
    \thisfloatpagestyle{empty}
\end{figure}

\begin{figure}[hbt!]
    \centering
    \includegraphics{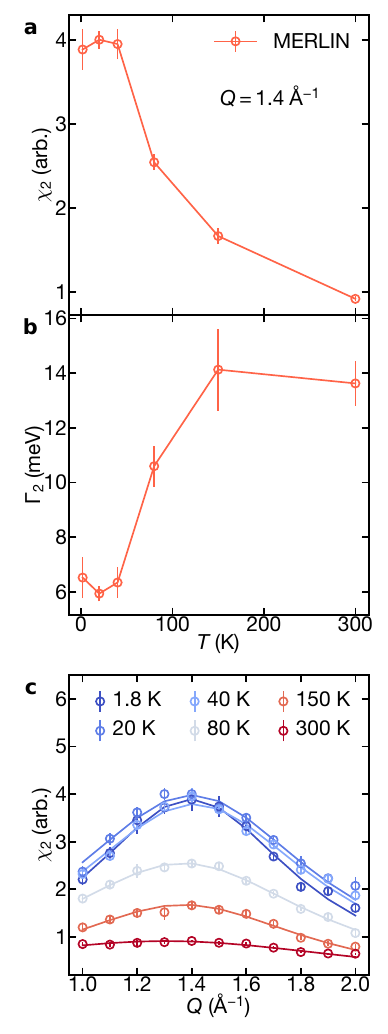}
    \caption{\textbf{a} The susceptibility, $\chi_2$, and \textbf{b} the relaxation rate, $\Gamma_2$, of fluctuations in the spin-liquid phase as a function of temperature at $1.35 \le Q \le 1.45$~\AA$^{-1}$, as determined from the INS data obtained on MERLIN. \textbf{c} The susceptibility, $\chi_2$, from the spin-liquid phase as a function of $Q$ at $T = 1.8,\;20,\;40,\;80,\;150,$ and $300$~K obtained from the MERLIN data.}
    \label{fig:merlin_metrics}
\end{figure}

\noindent The $S(Q,\Delta E)$ of \ce{ZnV2O4}  measured on the MERLIN thermal spectrometer at $T=1.8$~K is shown in Fig.~\ref{fig:merlin_slices}. This is composed of individual runs simultaneously measured at $E_i \in \{30,60,180\}$~meV and combined together using scale-factors derived from a vanadium standard measured at the same settings.  The data clearly show the same high-energy spin-fluctuations centred at $Q\approx1.4$~\AA$^{-1}$ as was seen in the PANTHER experiment (Fig. \ref{fig:INS}), extending all the way up to the edge of the kinematic range of MERLIN $\approx 40$~meV at $Q\approx1.4$~\AA$^{-1}$.  Examples of modelling of constant-$Q$ cuts at $0.95 \le Q \le 1.05$,~$1.35 \le Q \le 1.45$~and~$1.95 \le Q \le 2.05$~\AA$^{-1}$ of the MERLIN data at 1.5~K and 150~K are presented in Fig.~\ref{fig:INS_M}, with the data are modelled according to Equation~\ref{eq:Scalc}. Extracted fit parameters accounting only  for the high-energy response, $\chi''_2$, are plotted in Fig.~\ref{fig:merlin_metrics}.  These are qualitatively identical to those extracted from PANTHER, with only a slight increase in the values of $\Gamma_2(T)$.\\

\noindent The agreement between the modelled scattering on MERLIN and PANTHER is a good indication that sources of systematic error, including (non-exhaustive) instrumental calibration, background scattering and phonon scattering, are well accounted for in both the MERLIN and PANTHER experiments.

\end{appendices}

\end{document}